\documentclass[journal,twoside,10pt]{IEEEtran}
\usepackage[caption=false,font=normalsize,labelfont=sf,textfont=sf]{subfig}
\usepackage{amsmath,amsthm,amsfonts,amssymb,empheq}

\usepackage{array,soul,xcolor}
\usepackage{mathtools,cases,relsize,hyperref}
\hypersetup{
     colorlinks   = true,
     citecolor    = blue,
     linkcolor    = blue,
     urlcolor     = black,
}
\usepackage[T1]{fontenc}

\DeclareSymbolFont{myletters}{OML}{ztmcm}{m}{it}
\DeclareMathSymbol{\lamda}{\mathord}{myletters}{"15}

\usepackage{textcomp,xpatch}
\usepackage{stfloats,subfiles}
\usepackage{multirow}
\usepackage{graphicx}
\usepackage{url,verbatim}
\usepackage{cite,balance,hyperref}
\usepackage{nicefrac, xfrac}
\usepackage{aligned-overset}
\usepackage{bbding}
\usepackage{pifont}
\usepackage{wasysym}

\newtheoremstyle{colon}%
{}
{}
{\itshape}
{}
{\bfseries}
{:}
{ }
{}
\theoremstyle{colon}
\makeatletter
\xpatchcmd{\proof}{\@addpunct{.}}{\@addpunct{:}}{}{}
\makeatother
\makeatletter
\xpatchcmd{\proof}{\@addpunct{.}}{\@addpunct{:}}{}{}
\makeatother

\newtheorem{theorem}{Theorem}

\newtheorem{remark}{Remark}
\newtheorem{lemma}{Lemma}
\usepackage[ruled,vlined]{algorithm2e}
\include{pythonlisting}
\SetKwInput{KwRep}{Repeat}
\SetKwInput{KwUnt}{Until}

\newcommand{\src}{\mathrm{S}}
\newcommand{\relay}{\mathrm{R}}
\newcommand{\will}{\mathrm{W}}
\newcommand{\bob}{\mathrm{B}}
\newcommand{\eva}{\mathrm{E}}
\newcommand{\tom}{\mathrm{T}}
\newcommand{\dep}{p_{\rm dep}}

\newcommand{\sop}{p_{\rm sop}}
\renewcommand{\sec}{\calR_s}

\newcommand{\calR}{\mathcal{R}}

\newcommand{\calH}{\mathcal{H}}

\newcommand{\calF}{\mathcal{F}}
\newcommand{\calW}{\mathcal{W}}
\newcommand{\cald}{\textsf{D}}
\newcommand{\calEFoxH}{\calH^{1,0}_{0,1} = \big[z\big| \begin{smallmatrix} -  \\ (0,1)  \end{smallmatrix} \big]}
\newcommand{\calHFoxH}{\calH^{1,0}_{1,1} = \big[ z\big| \begin{smallmatrix} (1,1) \\ (0,1)  \end{smallmatrix}  \big]}
\newcommand{\calFFoxH}{\calH^{1,1}_{1,2} = \big[ -z\big| \begin{smallmatrix}  (1-a,1) \\    (0;1-b,1) \end{smallmatrix}  \big]}

\allowdisplaybreaks 
 \usepackage{xr-hyper}
 
\usepackage{makecell}

\begin{document}

\title{\huge{On Performance of IoT Networks with Coordinated NOMA Transmission: Covert Monitoring and Information Decoding}}
    \author{Thai-Hoc Vu, Anh-Tu Le, Ngo Hoang Tu, Tan N. Nguyen, and Miroslav Voznak
     \thanks{Thai-Hoc Vu, Anh-Tu Le, and Miroslav Voznak are with the Faculty of Electrical Engineering and Computer Science, VSB-Technical University of Ostrava, 17. Listopadu 2172/15, 708 00, Ostrava, Czechia (e-mail: hoc.vu.thai@vsb.cz, tu.le.anh.st@vsb.cz, miroslav.voznak@vsb.cz). 
     Ngo Hoang Tu is with the Department of Electrical and Information Engineering, Seoul National University of Science and Technology, Seoul 01811, South Korea (e-mail: ngohoangtu@seoultech.ac.kr).
     Tan N. Nguyen (\textit{corresponding author}) is with the Advanced Intelligent Technology Research Group, Faculty of Electrical and Electronics Engineering, Ton Duc Thang University, Ho Chi Minh City 70000, Vietnam (e-mail: nguyennhattan@tdtu.edu.vn). 
     }  
      \thanks{This research is funded by the European Union (EU) within the REFRESH project – Research Excellence For REgion Sustainability and High-tech Industries ID No. CZ.10.03.01/00/22\_003/0000048 of the European Just Transition Fund,  and also supported by the Ministry of Education, Youth and Sports of the Czech Republic (MEYS CZ) within a Student Grant Competition in the VSB – Technical University of Ostrava under project ID No. SGS SP2025/013.}  
}
\maketitle

\markboth{IEEE Internet of Things Journal, 2025}{IEEE Internet of Things Journal, 2025}

\begin{abstract}
\boldmath
{This work investigates the covertness and security performance of Internet-of-Things (IoTs) networks under Rayleigh fading environments.
Specifically, a cellular source transmits covert information to cell-edge users with the assistance of an IoT master node, employing a coordinated direct and relay transmission strategy combined with non-orthogonal multiple access (NOMA). This approach not only enhances spectrum utilization but also generates friendly interference to complicate a warden's surveillance or an eavesdropper's decoding efforts.
From a covertness perspective, we derive exact closed-form expressions for the detection error probability (DEP) under arbitrary judgment thresholds.
We then identify the optimal judgment threshold for the worst-case scenario, at which the warden minimizes its DEP performance.
Accordingly, we determine the effective region for user power allocation (PA) in NOMA transmission that satisfies the DEP constraint.
From a security perspective, we derive analytical expressions for the secrecy outage probability under two eavesdropping strategies using selection combining and maximal ratio combining. 
Based on this analysis, we propose an adaptive PA scheme that maximizes covert rate while ensuring the quality-of-service (QoS) requirements of legitimate users, the system's minimum covertness requirements, and supporting successive interference cancellation (SIC) procedures.
Furthermore, we design an adaptive PA scheme that maximizes the secrecy rate while ensuring the QoS requirements of legitimate users and SIC conditions. Numerical results demonstrate the accuracy of the analytical framework, while the proposed optimization strategies effectively adjust PA coefficients to maximize either the covert rate or the secrecy rate.}

\end{abstract}

\section{Introduction}\label{sec1}

\subsection{Research Context}\label{Sect:1A}
\IEEEPARstart{R}{ecently}, the widespread adoption of Internet of Things (IoT) applications, such as smart cities, multisensory augmented/virtual reality, remote surgery, the tactile Internet, and mobile streaming media, has led to substantial increase in the number of intelligent terminals.
This surge has, in turn, resulted in significantly higher traffic demands and presents considerable challenges for the development of fifth-generation (5G) networks, including spectrum scarcity, limited device energy, a restricted number of connections, and security concerns \cite{IoT-1,IoT-2,IoT-4}.  
In this context, non-orthogonal multiple access (NOMA) has emerged as a promising technology for addressing spectrum scarcity by exploiting multi-user superimposed position in the power domain at the transmitter, along with successive interference cancellation (SIC) at the receiver. As a result, NOMA achieves higher spectral efficiency and capacity compared to its orthogonal multiple access counterparts \cite{NOMA-1,NOMA-2,NOMA-3}. Moreover, NOMA has demonstrated its effectiveness in enhancing coverage and communication quality-of-service (QoS), particularly when integrated with cooperative communication strategies. For example, proximate users can act as relaying nodes to forward decoded messages from distant users \cite{NOMA-Cooperative-1}. Multi-relay nodes capable of decoding or amplifying functions are deployed to forward NOMA messages to their destinations \cite{NOMA-Cooperative-2}. Another approach is coordinated direct and relay transmission (CDRT)-NOMA \cite{CDRTNOMA-1}, which has garnered increased attention due to its ability to provide higher spectrum utilization and more sustainable cell coverage compared to earlier models. 

\subsection{Comprehensive Literature Overview} \label{Sect:1B}

This subsection provides a comprehensive overview of the state-of-the-art research topics, from which we highlight the timely and essential need for the research presented in this work. The details are as follows. 

\textit{CDRT-NOMA Investigation:}
Several notable studies have investigated the benefits of NOMA-based CDRT networks. For instance, Nguyen \textit{et al.} \cite{CDRTNOMA-2} proposed a novel IoT-assisted NOMA-CDRT paradigm, in which an IoT master node facilitates communication between a cellular source and a cell-edge user by employing NOMA protocols to simultaneously transmit to its IoT receiver and the cell-edge user. This approach improves spectral efficiency compared to conventional CDRT-NOMA systems \cite{CDRTNOMA-1}.
Unlike \cite{CDRTNOMA-1} and \cite{CDRTNOMA-2}, the authors in \cite{CDRTNOMA-3} and \cite{CDRTNOMA-4} explored NOMA-CDRT with short-packet communications under imperfect SIC and co-channel interference, while also developing a dynamic relay selection strategy to enhance transmission reliability for both cell-edge and cell-center users.
In \cite{CDRTNOMA-5}, Jee \textit{et al.} proposed an incremental relaying strategy to improve the throughput of distant users while maintaining the requisite performance for proximate users.
Meanwhile, \cite{CDRTNOMA-6} analyzed the performance of decode-and-forward (DF)-assisted CDRT-NOMA networks under constraints imposed by the interference temperature limit caused by the primary network on the transmission power.

\textit{Covert Communication Investigation:}
On the other hand, with the increasing transmission of confidential and sensitive data (e.g., financial information, tactical intelligent networks, and identity verification) over the open wireless medium, ensuring secrecy and privacy has become crucial to the development of sixth-generation wireless technologies. This challenge necessitates the use of physical layer security (PLS), which mitigates information leakage by leveraging random noise and fading channels \cite{shiu2011physical}. Nevertheless, current PLS techniques are insufficient to fully protect the content of communication, especially when the transmission must remain undetectable. Consequently, covert communication has emerged as a novel security paradigm aimed at enabling wireless information transmission from a legitimate transmitter to an intended receiver while maintaining a minimal likelihood of detection by a warden.
In \cite{Corvert-1} and \cite{Corvert-2}, He \textit{et al.} analyzed the achievable covert rate in scenarios where the warden lacks accurate knowledge of the ambient noise level.
In contrast, Hu \textit{et al.} \cite{Corvert-3} investigated channel inversion power control assisted by a full-duplex receiver to eliminate the need for feedback from the transmitter to the receiver, thereby concealing the transmitter's presence from the warden and enabling covert transmission.
Focusing on untrusted relaying networks, Forouzesh \textit{et al.} \cite{Corvert-4} developed a sequential convex approximation method to optimize the secrecy rate in the presence of multiple wardens.
In \cite{TuWCL2024}, Le \textit{et al.} analyzed the performance of PLS and covert communication in a power-frequency multiple-access system by deriving expressions for the connection outage probability, secrecy outage probability (SOP), and detection error probability (DEP).
They also established an approximation for the transmit signal-to-noise ratio (SNR) and the warden's detection threshold to minimize DEP.
Moreover, Gao \textit{et al.} \cite{Corvert-5} studied the DEP of the warden and covert capacity maximization by employing relay selection strategies, including random selection and superior-link selection.

\textit{Conventional NOMA Covertness Investigation:}
Due to the superior capabilities of NOMA in encoding multi-user communication in the power domain, which inherently offers a novel steganographic approach to concealing sensitive information, there has been increasing interest in exploring covert communication within NOMA systems. 
The initial integration of covert communication with NOMA was introduced by Tao \textit{et al.} \cite{tao2020covert}, who focused on deriving the exact expression for DEP and determining the optimal power allocation (PA) strategy to maximize the effective covert rate.
In \cite{Jiang2020Feb}, Jiang \textit{et al.} examined covert performance in device-to-device communication by utilizing cooperative power-domain NOMA at the base station to decode the covert signal.
Duan \textit{et al.} \cite{duan2023covert} investigated the optimal PA to maximize the expected covert rate, subject to DEP and COP constraints under channel distribution information in uplink NOMA systems with covert communication.
In contrast to \cite{duan2023covert}, Li \textit{et al.} \cite{li2023covert} studied covert communication alongside secure transmission in downlink NOMA networks under the condition of imperfect SIC. They proposed an analytical framework to maximize the effective covert rate while maintaining constraints related to covertness, secrecy, and reliability.
Focusing on active reconfigurable repeater-aided NOMA networks, Le \textit{et al.} \cite{TuIOT2024} derived closed-form expressions for outage probability, SOP, and DEP, and proposed two optimization strategies to ensure user outage fairness.
\textit{CDRT-NOMA with PLS and Research Gap in Covertness:}
To the best of the authors' knowledge, only a few studies \cite{Lei2022Secure,Chen2017PLS,Pei2019Secure,Lv2020Secure,Xu2024Secure} have investigated PLS in the context of CDRT-NOMA. 
To be specific, Lei \textit{et al.} \cite{Lei2022Secure} derived a closed-form expression for the ergodic secrecy sum rate and an asymptotic expression for the SOP within a synergy of  CDRT-NOMA and physical-layer network coding strategies.
The frameworks in \cite{Chen2017PLS} and \cite{Pei2019Secure} focused on an artificial noise design to simultaneously disrupt eavesdroppers while preserving communication quality for legitimate users in CDRT-NOMA systems. 
Meanwhile, Lv \textit{et al.} \cite{Lv2020Secure} analyzed the secrecy performance of an untrusted relay-based adaptive cooperative jamming strategy for both downlink and uplink NOMA transmissions.
In \cite{Xu2024Secure}, Xu \textit{et al.} derived closed-form expressions for the ergodic secrecy rate and further introduced a joint optimal PA and interference control strategy.
Nevertheless, covert communication in CDRT-NOMA networks remains  a notably unexplored research area. Fundamentally, the CDRT-NOMA strategy offers two flexible and efficient mechanisms for concealing information.
In the first phase, the use of NOMA allows the covert message to be masked by an overt one, thereby providing a shielding effect. In the second phase, the simultaneous transmission of information from the source to proximate users, while the relay forwards data to distant users, creates an additional layer of protection for covert communication, effectively acting as a form of free jamming that efficiently exploits spectral resources. 

\begin{table*}[!th]
    \renewcommand{\arraystretch}{1.2}
    \centering
    \caption{Novelty comparison of the proposed study and state-of-the-art related works.}    
    \label{tab:compairation}
    \begin{tabular}{|l|c|c|c|c|c|c|c|c|c|c|c|c|c|c|}
    \hline
    \multirow{2}{*}{\hspace{0.9cm} \textbf{Contexts}} & \multicolumn{9}{c|}{\textbf{State-of-the-art related studies}} & \multirow{2}{*}{\textbf{Our work}}
    \\ \cline{2-10}
        {}     & \cite{CDRTNOMA-1,CDRTNOMA-2,CDRTNOMA-3,CDRTNOMA-4,CDRTNOMA-5,CDRTNOMA-6} & \cite{Corvert-1,Corvert-2} & \cite{Corvert-3} & \cite{Corvert-4} & \cite{TuWCL2024} & \cite{Corvert-5,tao2020covert,Jiang2020Feb,duan2023covert} & \cite{li2023covert} & \cite{TuIOT2024} & \cite{Lei2022Secure,Chen2017PLS,Pei2019Secure,Lv2020Secure,Xu2024Secure} & {} \\ \hline
    \hline
    CDRT-NOMA networks        & \Checkmark & X & X & X & X & X & X & X & \Checkmark & \Checkmark \\ \hline
    Covert communication      & X & \Checkmark & \Checkmark & \Checkmark & \Checkmark & \Checkmark & \Checkmark & \Checkmark & X & \Checkmark \\ \hline
    Physical layer security   & X & \Checkmark & \Checkmark & X & \Checkmark & X & \Checkmark & \Checkmark & \Checkmark & \Checkmark \\ \hline
    DEP analysis              & X & X & \Checkmark & \Checkmark & \Checkmark & \Checkmark & \Checkmark & \Checkmark & X & \Checkmark \\ \hline
    SOP analysis              & X & X & X & X & \Checkmark & X & X & \Checkmark & \Checkmark & \Checkmark \\ \hline
    Covert rate maximization  & X & \Checkmark & \Checkmark & X & X & \Checkmark & \Checkmark & X & X & \Checkmark \\ \hline
    Secrecy rate maximization & X & X & X & \Checkmark & X & X & X & X & X & \Checkmark \\ \hline

    \hline
    
    \end{tabular}
\end{table*}

\subsection{Novelty and Contributions}
To fulfill the research gap discussed in Section~\ref{Sect:1B}, this study synergistically investigates the performance of both covert communication and PLS within CDRT-NOMA networks.
The novelty of this work compared with state-of-the-art studies is summarized in Table~\ref{tab:compairation}, whereas our main contributions are discussed in more detail below.

    1) \textbf{Novel Framework}: This study is the first to comprehensively investigate the synergy between covertness communication and PLS in transmission from a source to cell-edge users within CDRT-NOMA shared cellular and IoT networks under Rayleigh fading channels. 
    Unlike prior works \cite{Lei2022Secure,Chen2017PLS,Pei2019Secure,Lv2020Secure,Xu2024Secure}, which investigate standalone PLS within conventional CDRT-NOMA systems \cite{CDRTNOMA-1}, this work adopts a more flexible CDRT-NOMA framework \cite{CDRTNOMA-2}, aiming to improve spectrum utilization for diverse IoT paradigms by jointly addressing both covert communication and PLS. Specifically, we consider a scenario in which a source node transmits a covert signal to a cell-edge user with the assistance of an IoT master node across two communication phases. These covert transmissions are subject to monitoring by a potential user, who may act either as a warden (i.e., attempting covert signal detection) or as an eavesdropper (i.e., attempting to decode the covert information). 
    Furthermore, it is worth mentioning that none of the studies in \cite{Lei2022Secure,Chen2017PLS,Pei2019Secure,Lv2020Secure,Xu2024Secure} investigated covert rate and secrecy rate maximization problems under specific network constraints, which are both practical and significant challenges in secure and covert communications.
    This work aims to address these critical issues in greater depth.
    
    2) \textbf{Covertness Perspective}: In the presence of potential warden surveillance, we derive a closed-form expression for the DEP of the warden.
    We then obtain the exact optimal judgment threshold that minimizes the DEP in two-phase communications. 
    Based on this worse-case scenario, we identify the effective range of PA coefficients in NOMA transmissions that guarantees the minimum level of covertness, as defined by the DEP threshold, for covert information transmission.    
    
    3) \textbf{Security Perspective}: We present preliminary analysis and derive exact analytical expressions for the SOP under two scenarios where the eavesdropper uses either selection combining (SC) or maximal ratio combining (MRC) diversity techniques. 

    4) \textbf{Resource Allocation for Covert Rate Maximization}: We formulate a PA optimization problem aimed at maximizing the covert rate, while satisfying the QoS requirements of legitimate users, ensuring SIC operation in NOMA protocols, and preventing covert monitoring under a specified covertness constraint.
    This non-convex PA optimization problem is addressed through a convex approximation transformation, incorporating a fairness condition.

    5) \textbf{Resource Allocation for Secrecy Rate Maximization}: Additionally, we formulate a separate PA optimization problem that maximizes the secrecy rate under either SC or MRC eavesdropping scenarios.
    This formulation satisfies the QoS requirements of legitimate users and ensures SIC operation.
    The non-convex problem is reformulated into a convex approximation that can be solved using standard convex solvers, such as CVX. An iterative optimization algorithm is then proposed to obtain a suboptimal yet efficient solution.
    
    6) \textbf{Validation and Insights}: All analytical and optimization results are validated through Monte Carlo simulations, from which several key insights are observed, as summarized below. 
    \begin{itemize}
        \item \textit{DEP Trends}: The analysis reveals that, from a system design perspective, implementing the CDRT-NOMA strategy during the second communication phase provides more effective protection for covert messages compared to the first phase. Moreover, an eavesdropper can significantly reduce the DEP by increasing the number of monitoring sample tests used for covert signal classification.
        
        \item \textit{SOP Trends}: Quantitative analysis of SOP performance under different eavesdropping scenarios shows that the system performs better when the eavesdropper uses SC rather than MRC. Notably, an optimal transmission power range between 12 dBm and 14 dBm is identified, within which the SOP is minimized.  
        
       \item \textit{Covert Rate Trends under Max-Min Fairness Optimization}: Evaluation of covert rate performance shows that as the covertness requirement (i.e., DEP target) increases from 0.7 to 0.9, the system must increase the source transmit power by at least 5 dB to maintain equivalent covert rate performance.
       For a given transmit power, the impact of covertness requirements on the covert rate is significant only when the IoT user's target data rate is below 1 bps/Hz; beyond this threshold, the effect becomes negligible.

       \item \textit{Secrecy Rate Maximization}: Assessments of secrecy rate maximization demonstrate that the proposed approach converges rapidly within five iterations. The system maintains stable performance when the IoT user's target data rate is below 0.75 bps/Hz, but the secrecy rate sharply degrades to zero when the target data rate exceeds 1.5 bps/Hz.
       
    \end{itemize}

\begin{figure}[!t]
    \centering   
    \includegraphics[width=\linewidth]{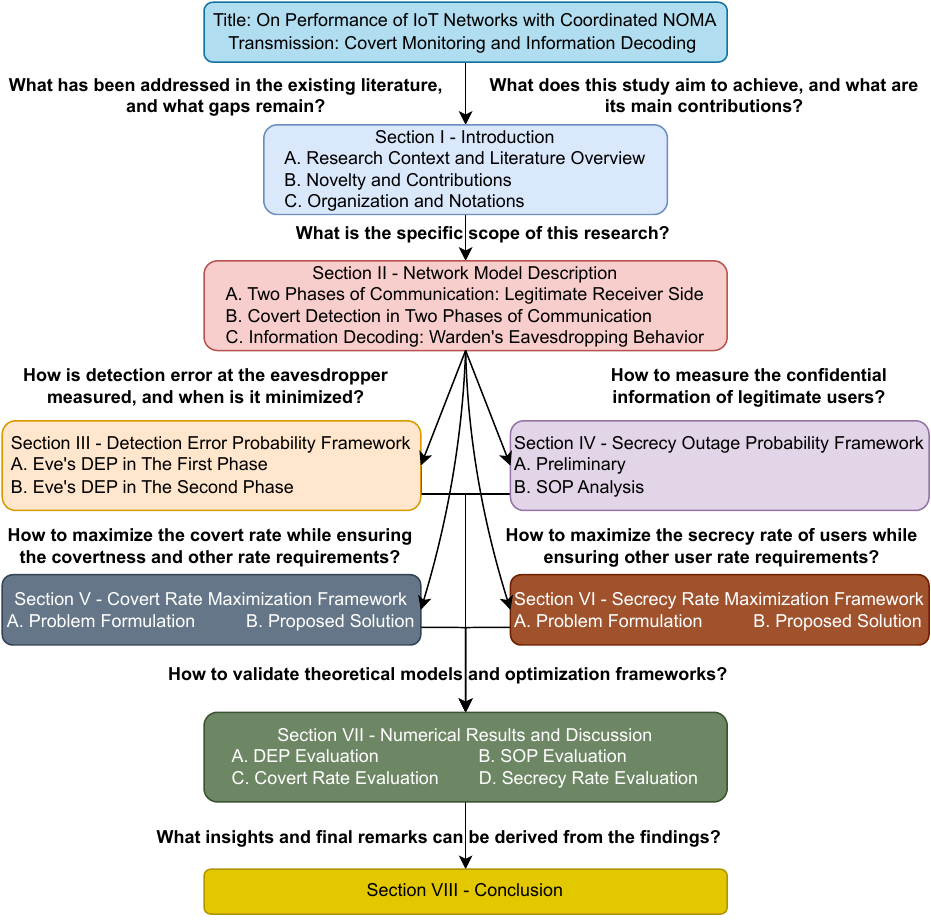}
    \caption{Outline of the paper.}
    \label{fig:Outline}
\end{figure}

\subsection{Organization and Notations}
The remainder of this paper is organized as follows. 
Section~\ref{sec2} describes the proposed system model and communication analysis.
Section~\ref{sec3} presents a DEP framework, including the derivation of DEP expressions, the optimal judgment threshold for minimizing the DEP, and the effective PA region required to satisfy the minimum covertness threshold.
Section~\ref{sec4} derives closed-form expression for the SOP under both SC and MRC eavesdropping scenarios. 
Section~\ref{sec5} and Section~\ref{sec6} introduce adaptive PA solutions aimed at maximizing the covert rate and secrecy rate, respectively, under several practical system constraints.
Section~\ref{sec7} provides numerical results based on Monte Carlo simulations to validate the proposed analytical and optimization frameworks. 
Finally, Section~\ref{sec8} concludes the paper. 
A flowchart of our research methodology is presented in Fig.~\ref{fig:Outline}.

\textit{Mathematical Notations}:  
$\Pr(\bullet)$ denotes the probability operator. 
$F_X(\bullet)$ and
$f_X(\bullet)$ represent the cumulative distribution function (CDF) and the probability density function (PDF) of a random variable $X$, respectively. 
$\mathbb{E}(\bullet)$ and $\mathbb{V}(\bullet)$ represent the expectation and variance operators, respectively.
\([x]^+ = \max\{x,0\}\) denotes
the positive part function, whereas
$\partial f/\partial x$ denotes the first-order partial derivative of function $f$ with respect to $x$.
Furthermore, \(\gamma(\bullet,\bullet) \) and \(\Gamma(\bullet,\bullet)\) represent the lower and upper incomplete Gamma functions \cite[Eqs. (8.350.1) and (8.350.2)]{Gradshteyn2014}, respectively.
\({_{1}\calF_{1}}(\bullet;\bullet;\bullet)\) denotes the degenerate (confluent) hypergeometric function \cite[Eq. (9.21)]{Gradshteyn2014}.  
Finally, \(\calW(\bullet)\), \(\calH(\bullet)\),
and
\(\calH^{\cdot,\cdot}_{\cdot,\cdot}\left[\begin{smallmatrix} \dotsc\\ \dotsc \end{smallmatrix}\left| \bullet \right. \right]\)
represent the Lambert~W function \cite[Eq. (1.5)]{Corless1996},
the Heaviside step function \cite{Gradshteyn2014},
and
the Fox-H function \cite[Eq. (1.1.1)]{Prudnikov1986}, respectively.

\section{Network Model Description}\label{sec2}
\begin{figure} [t!]
    \centering
    \includegraphics[width=0.7\columnwidth]{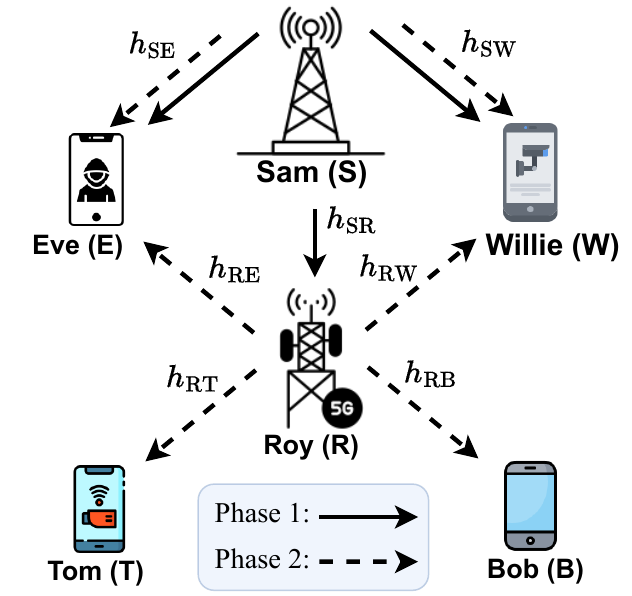}
    \caption{A synergistic ecosystem of PLS and covert communication in a downlink CDRT-NOMA IoT network.}
    \label{fig:1}
\end{figure}

Fig.~\ref{fig:1} illustrates a synergistic ecosystem combining PLS and covert communication within a downlink CDRT-NOMA IoT network. We consider that a primary system consists of a source, referred to as Sam ($\src$), a distant user, referred to as Bob ($\bob$), and a proximate user, referred to as Willie ($\will$), while a secondary system comprises an IoT master node, refereed to as Roy ($\relay$), and an IoT receiver, referred to as Tom ($\tom$). Sam intends to send a covert task to Bob using NOMA encoding as a shield but is unable to reach Bob due to limited communication distance, severe shadowing, or blockages. Meanwhile, Roy, who seeks to communicate with Tom but encounters spectrum limitations, can assist Sam by  facilitating cooperative communication to help complete Sam's information exchange with Bob. In this work, we focus on the PLS performance of the primary system, where a potential user, denoted Eve ($\eva$), acts either as a warden monitoring Bob's covert information or as an eavesdropper with strong extraction and classification capabilities to wiretap Bob's transmissions. 
Moreover, it is assumed that perfect channel state information (CSI) is available at all network nodes \cite{Lei2022Secure,Chen2017PLS,Pei2019Secure,Lv2020Secure,Xu2024Secure,vu2025study,tu2024irs,Tu2021performance,tu2023short}.

To enhance Bob's covert transmission while efficiently utilizing the spectrum, we adopt a flexible CDRT-NOMA strategy \cite{CDRTNOMA-2}, in which Sam and Roy make the shield for Bob's covert signal by allocating higher NOMA power levels to the public signals of Willie in the first phase and those of Tom in the second phase, respectively.
Additionally, during Roy’s NOMA broadcasting in the second phase, Sam simultaneously transmits a new signal to Willie instead of maintaining silence.
This transmission also acts as a jamming signal to disrupt Eve's covert monitoring. 
However, Roy's broadcasting interferes with Willie's signal reception. To counteract this covert monitoring and enhance overall system performance, it is essential to investigate the PA policy for two-phase NOMA communications and the joint transmit power resources. 
On the other hand, it is worth mentioning that when Sam intends to transmit a covert signal to Willie, the CDRT-NOMA strategy remains applicable with a slightly modified setup. In this case, Sam protects Willie's covert signal by assigning higher NOMA power levels to the public signals of Bob in the first phase, and the same approach is adopted by Roy in the second phase. To support the resilience of Willie's signal, Sam retransmits the signal from the previous phase to Willie, who can then reuse the decoded information from Bob to partially mitigate the interference caused by Roy's broadcasting. Nevertheless, since the same performance analysis method applies to both scenarios, we omit detailed consideration of the case in which Sam sends a covert signal to Willie.

In this study, we assume that all nodes in the network operate in half-duplex mode and are equipped with a single antenna. For the wireless operating environment, we consider the impact of Rayleigh fading on the channel \(h_{\textsf{XY}}\) for links between \(\textsf{X}\in\{\src,\relay\}\) and \(\textsf{Y}\in\{\relay,\will,\eva,\tom,\bob\}\).
The PDF and CDF of $|h_{\textsf{XY}}|^2$ are given, respectively, as \cite{tu2024irs,Tu2021performance,tu2023short}
\begin{align} \label{eq:PDF:hX}
    f_{|h_{\textsf{XY}}|^2}(x) &= \exp(-x/\lambda_\textsf{XY})/\lambda_{\textsf{XY}}, \\ \label{eq:CDF:hX}
    F_{|h_{\textsf{XY}}|^2}(x) &= 1 - \exp(-x/\lambda_\textsf{XY}),
\end{align}
where $\lambda_{\textsf{XY}}$ denotes the average channel gain. Moreover, the noise at the receiving node \(\textsf{Y}\) is modeled as additive white Gaussian noise (AWGN) with zero mean and variance \(\sigma^2\).

\subsection{Two Phases of Communication: Legitimate Receiver Side}
\subsubsection{First-Phase Communication}
In the first phase of communication, Sam employs the NOMA protocol to simultaneously transmit a public signal \(s_\will\) intended for Willie and a covert signal \(c_\bob\) intended for Bob, using the composite signal \(x_1 = \sqrt{\alpha_\will}s_\will + \sqrt{\alpha_\bob}c_\bob\),
satisfying
\(\mathbb{E}\{|s_\will|\}=\mathbb{E}\{|c_\bob|\} = 0 \), \(\mathbb{E}\{|s_\will|^2\}=\mathbb{E}\{|c_\bob|^2\} = P_s\), \(\alpha_\will > \alpha_\bob\), and \(\alpha_\will + \alpha_\bob =1\), 
with \(\alpha_\will\) and \(\alpha_\bob\) representing the PA coefficients of \(s_\will\) and \(c_\bob\), respectively.

Under Sam's broadcast, the signals received by Roy and Willie can be expressed, respectively, as
\begin{align}
    y_\relay[1] &=  (\sqrt{\alpha_\will}s_\will + \sqrt{\alpha_\bob}c_\bob)h_{\src\relay} + n_\relay,\\
    y_\will[1] &=  (\sqrt{\alpha_\will}s_\will + \sqrt{\alpha_\bob}c_\bob)h_{\src\will} + n_\will,
\end{align}
where \(n_\relay\sim\mathcal{CN}(0,\sigma^2)\) and \(n_\will\sim\mathcal{CN}(0,\sigma^2)\) denote the AWGN signals at Roy and Willie, respectively. 

Based on the two expressions above, the signal-to-interference-plus-noise ratios (SINRs) for decoding \(s_\will\) and \(c_\bob\) at Roy (with \(s_\will\) eliminated via SIC), as well as for decoding \(s_\will\) at Willie, can be expressed in a manner similar to that presented in \cite{Lei2022Secure, Chen2017PLS, Pei2019Secure, Lv2020Secure, Xu2024Secure}, as follows:
\begin{align}
\label{Roy-SNR-1-1}
    \gamma_{\relay}^{s_\will} &= \frac{\alpha_\will P_s |h_{\src\relay}|^2}{\alpha_\bob P_s |h_{\src\relay}|^2  + \sigma^2}, \,
    \gamma_{\relay}^{c_\bob} = \frac{\alpha_\bob P_s |h_{\src\relay}|^2}{\sigma^2},\\
 \label{Will-SNR-1-1}
    \gamma_{\will}^{s_\will} &= \frac{\alpha_\will P_s |h_{\src\will}|^2}{\alpha_\bob P_s |h_{\src\will}|^2  + \sigma^2}.
\end{align}

\subsubsection{Second-Phase Communication}
During this phase, Roy employs NOMA to encode a public signal \(s_\tom\) intended for Tom and the previously decoded covert signal \(c_\bob\) intended for Bob, using the composite signal \(x_2 = \sqrt{\beta_{\tom}}s_\tom + \sqrt{\beta_{\bob}}c_\bob\),
where \(\beta_{\tom}\) and \(\beta_{\bob}\) represent the power allocation ratios for \(s_\tom\) and \(c_\bob\), respectively. 
The signals are defined such that
\(\mathbb{E}\{|s_\tom|\}=\mathbb{E}\{|c_\bob|\} = 0 \), \(\mathbb{E}\{|s_\tom|^2\}=\mathbb{E}\{|c_\bob|^2\} = P_r\), \(\beta_{\bob} < \beta_{\tom}\),
and \(\beta_{\bob} + \beta_{\tom} = 1\).
Accordingly, the signals received by Bob and Tom can be expressed, respectively, as
\begin{align}
    y_\bob[2] &=  (\sqrt{\beta_{\tom}}s_\tom + \sqrt{\beta_{\bob}}c_\bob)h_{\relay\bob} + n_\bob,\\
     y_\tom[2] &=  (\sqrt{\beta_{\tom}}s_\tom + \sqrt{\beta_{\bob}}c_\bob)h_{\relay\tom} + n_\tom,
\end{align}
where \(n_\bob\sim\mathcal{CN}(0,\sigma^2)\) and \(n_\tom\sim\mathcal{CN}(0,\sigma^2)\) denote the AWGN signals at Bob and Tom, respectively. 

Based on these signals, the SINRs for decoding \(s_\tom\) and \(c_\bob\) at Bob (with \(s_\tom\) eliminated via SIC), and for decoding \(s_\tom\) at Tom, are given, respectively, by
\begin{align}
\label{Bob-SNR-1-1}
    \gamma_{\bob}^{s_\tom} &= \frac{\beta_{\tom} P_r |h_{\relay\bob}|^2}{\beta_{\bob} P_r |h_{\relay\bob}|^2  + \sigma^2}, \,
    \gamma_{\bob}^{c_\bob} = \frac{\beta_{\bob} P_r |h_{\relay\bob}|^2}{\sigma^2},\\
\label{Tom-SNR-1-1}
    \gamma_{\tom}^{s_\tom} &= \frac{\beta_{\tom} P_r |h_{\relay\tom}|^2}{\beta_{\bob} P_r |h_{\relay\tom}|^2  + \sigma^2}.
\end{align}
Meanwhile, the signal received by Willie during this phase is given by
\begin{align}
    y_\will[2] &=  (\sqrt{\beta_{\tom}}s_\tom + \sqrt{\beta_{\bob}}c_\bob)h_{\relay\will} + \widehat{s}_\will h_{\src\will} + n_\will,
\end{align}
where \(\widehat{s}_\will\) is a new signal transmitted by Sam to Willie, such that \(\mathbb{E}\{|\widehat{s}_\will|\} = 0 \)
and
\(\mathbb{E}\{|\widehat{s}_\will|^2\} = P_s\). 
Based on this, the SINR for decoding \(\widehat{s}_\will\) at Willie is given by
\begin{align}
\label{Will-SNR-2}
    \gamma_{\will}^{\widehat{s}_\will} = \frac{P_s |h_{\src\will}|^2}{P_r |h_{\relay\will}|^2  + \sigma^2}.
\end{align}

\subsection{Covert Detection in Two Phases of Communication}

Eve aims to determine whether Bob's secret information is being transmitted based on her observed signal. Specifically, Eve employs a binary hypothesis test, with \(\calH_1\) representing the presence of the covert signal \(c_\bob\), and \(\calH_0\) representing the absence of  \(c_\bob\), to distinguish her received signal.

\subsubsection{Covert Detection in First-Phase Communication}
During this phase, the signal received by Eve for covert monitoring can be expressed in a manner similar to that presented in \cite{shiu2011physical,Corvert-1,Corvert-2,Corvert-3,Corvert-4,TuWCL2024,Corvert-5}, as follows:
\begin{align}
    y_\eva[1] = 
    \begin{cases}
        (\sqrt{\alpha_\will}s_\will + \sqrt{\alpha_\bob}c_\bob)h_{\src\eva} + n_\eva, & \calH_1, \\
        \sqrt{\alpha_\will}s_\will h_{\src\eva} + n_\eva, & \calH_0,
    \end{cases}
\end{align}
where \(n_\eva\sim\mathcal{CN}(0,\sigma^2)\) denotes the AWGN signal at Eve. 

Using a radiometer on \(N\) signal samples, Eve applies the Nyeman--Pearson criterion for the detection of covert signals.
The decision rule is given by \cite{li2023covert}
\begin{align}
    \Phi_{\eva}[1] = \frac{1}{N} \sum_{n=1}^N|y_{\eva}^n[1]|^2 \overset{\cald_1}{\underset{\cald_0}{\gtrless}} \omega_1,
\end{align}
where \(\cald_1\) and \(\cald_0\) denote decisions in favor of \(\calH_1\) and \(\calH_0\), respectively, and  \(\omega_1\) is the judgment threshold. For a finite number of samples, as discussed in \cite{8761935}, the test statistic converges to the following expression: 
\begin{align}
   \Phi_{\eva}[1] 
   \to  \mathbb{E}\left\{  \sum_{n=1}^N\frac{|y_{\eva}^n[1]|^2 }{N}\right\} = \begin{cases}
        P_s \tau_{\src\eva} + N\sigma^2, & \calH_1, \\
        \alpha_\will P_s \tau_{\src\eva} + N\sigma^2, & \calH_0,
    \end{cases}\hspace{-3pt}
\end{align}
where \(\tau_{\src\eva}\triangleq \sum_{n=1}^N|h^n_{\src\eva}|^2\).

From an operational perspective, Eve must determine whether the received signal contains the covert message.
A missed detection event occurs when the covert signal \(c_\bob\) is transmitted, i.e., under hypothesis $\calH_1$, but the decision favors $\cald_0$. The missed detection probability is given by 
\begin{align}
    p_m[1] &= \Pr\left( \Phi_{\eva}[1], \cald_1| \calH_0 \right)= \Pr\left( \Phi_{\eva}> \omega_1 | \calH_0 \right)\nonumber\\
    &= \Pr(\alpha_\will P_s \tau_{\src\eva} + N\sigma^2 > N\omega_1).
\end{align}
Conversely, a false alarm occurs when no covert signal \(c_\bob\) is transmitted, i.e., under hypothesis $\calH_0$, but the decision favors $\cald_1$, with probability
\begin{align}
    p_f[1] &= \Pr\left( \Phi_{\eva}[1], \cald_0| \calH_1 \right)= \Pr\left( \Phi_{\eva}< \omega_1 | \calH_1 \right)\nonumber\\
    &= \Pr(P_s \tau_{\src\eva} + N\sigma^2 < N\omega_1).
\end{align}
Assuming equal prior probabilities for \(\calH_0\) and \(\calH_1\), Eve’s overall detection performance is evaluated by its DEP, defined as
\(\dep[1]= p_m[1] + p_f[1] \) \cite{TuIOT2024}.
Intuitively, Eve seeks to minimize this DEP $\dep[1]$ by optimizing the judgment threshold \(\omega_1\), which can be mathematically expressed as 
\begin{align}
\label{dep:scheme1-1}
    \min_{\omega_1} \; \dep[1]\; \text{s.t.}\; \omega_1 \geq 0.
\end{align}

\subsubsection{Covert Detection in Second-Phase Communication}
In this phase, Eve also applies a binary hypothesis test similar to that of the first phase.
The signal received by Eve for covert monitoring in the second phase can be derived in a manner similar to that presented in \cite{shiu2011physical,Corvert-1,Corvert-2,Corvert-3,Corvert-4,TuWCL2024,Corvert-5}, as follows:
\begin{align}
    y_\eva[2] = 
    \begin{cases}
        (\sqrt{\beta_{\tom}}s_\tom + \sqrt{\beta_{\bob}}c_\bob)h_{\relay\eva} + \widehat{s}_\will h_{\src\eva} +  n_\eva, \hfill \calH_1, \\
        \sqrt{\beta_{\tom}}s_\tom h_{\relay\eva} + s_\will h_{\src\eva} +  n_\eva, \hfill \calH_0.
    \end{cases}
\end{align}

For an \(N\)-sample signal collection with the judgment threshold \(\omega_2\), Eve seeks to minimize its DEP in the second phase, denoted by $\dep[2]$, by optimizing $\omega_2$ such that
\begin{align}
\label{dep:scheme1-2}
    \min_{\omega_2} \; \dep[2] \; \text{s.t.}\; \omega_2 \geq 0,
\end{align}
where $\dep[2] = p_m[2] + p_f[2]$, with $p_m[2]$ and $p_f[2]$ being the missed detection and false alarm probabilities, respectively. Upon the statistic test of
\begin{align}
   \Phi_{\eva}[2] 
   &\to  \mathbb{E}\left\{  \sum_{n=1}^N\frac{|y_{\eva}^n[2]|^2 }{N}\right\} \nonumber\\
   &= \begin{cases}
        P_r \tau_{\relay\eva} + P_s\tau_{\src\eva} + N\sigma^2, & \calH_1, \\
        \beta_{\tom} P_r \tau_{\relay\eva} + P_s\tau_{\src\eva} + N\sigma^2, & \calH_0,
    \end{cases}
\end{align}
with \(\tau_{\relay\eva}\triangleq \sum_{n=1}^N|h^n_{\relay\eva}|^2\), 
$p_m[2]$ and $p_f[2]$ can be calculated, respectively, as
\begin{align}
    p_m[2] &= \Pr\left( \Phi_{\eva}[2], \cald_1| \calH_0 \right)= \Pr\left( \Phi_{\eva}[2]> \omega_2 | \calH_0 \right)\nonumber\\
    &= \Pr(\beta_{\tom} P_r \tau_{\relay\eva} + P_s\tau_{\src\eva} + N\sigma^2 > N\omega_2), \\
    p_f[2] &= \Pr\left( \Phi_{\eva}[2], \cald_0| \calH_1 \right)= \Pr\left( \Phi_{\eva}< \omega_1 | \calH_1 \right)\nonumber\\
    &= \Pr(P_r \tau_{\relay\eva} + P_s\tau_{\src\eva} + N\sigma^2 < N\omega_2).
\end{align} 

\subsection{Information Decoding: Warden's Eavesdropping Behavior}
\subsubsection{First-Phase Communication}
As an eavesdropper, Eve possesses strong signal classification capabilities, allowing him to distinguish between the covert signal \(c_\bob\) and the public signal \(s_\will\).
Assuming that Eve successfully eliminates \(s_\will\) using SIC, the resulting SNR for detecting \(c_\bob\) in the first phase can be derived in a manner similar to that presented in \cite{Lei2022Secure,Chen2017PLS,Pei2019Secure,Lv2020Secure,Xu2024Secure}, as follows:
\begin{align}
\label{Eve-SNR-1-1}
    \gamma_{\eva}^{c_\bob}[1] = {\alpha_\bob P_s |h_{\src\eva}|^2}/{\sigma^2}.
\end{align}

\subsubsection{Second-Phase Communication}
Similarly, in the second phase, after eliminating the public signal \(s_\tom\) by SIC, Eve’s SINR for detecting \(c_\bob\) after eliminating \(s_\tom\) is given by
\begin{align}
\label{Eve-SNR-1-2}
    \gamma_{\eva}^{c_\bob}[2] = \frac{\beta_{\bob} P_r |h_{\relay\eva}|^2}{ P_s|h_{\src\eva}|^2 + \sigma^2}.
\end{align}

Given the SINRs in \eqref{Eve-SNR-1-1} and \eqref{Eve-SNR-1-2}, Eve can further enhance her decoding capability by applying either SC or MRC. The corresponding end-to-end SNR is expressed as
\begin{subequations}
  \label{Eve-SNR}
    \begin{empheq}[left={\gamma_{\eva} = \empheqlbrace\,}]{align}
      & \max\{\gamma_{\eva}^{c_\bob}[1],\gamma_{\eva}^{c_\bob}[2]\}, & \text{SC},
        \label{eq:SC} \\
      & \gamma_{\eva}^{c_\bob}[1] + \gamma_{\eva}^{c_\bob}[2], & \text{MRC}.
        \label{eq:MRC}
    \end{empheq}
\end{subequations}

%
\section{Detection Error Probability Framework}\label{sec3}

In this section, we analyze Eve's DEP across the two communication phases under arbitrary settings of the judgment threshold.
We then proceed to derive the optimal threshold that minimizes the DEP and provide guidance for determining a feasible PA policy in each phase to satisfy a minimum DEP protection requirement, denoted by \(\dep^{\rm th}\), which ensures the misclassifications of Eve’s detection attempts.

\subsection{Eve's DEP in The First Phase} To solve problem \eqref{dep:scheme1-1}, we first derive the DEP at Eve as
\begin{align}
\label{dep:close-1}
    \dep[1] 
    &= 1 - F_{\tau_{\src\eva}}\left( \frac{N(\omega_1 - \sigma^2)}{\alpha_\will P_s } \right) + F_{\tau_{\src\eva}}\left( \frac{N(\omega_1 - \sigma^2)}{P_s} \right)\nonumber\\
    &
    =\frac{\Gamma\left(N , u \right)}{\Gamma(N )} + 1 - \frac{\Gamma\left( N , v \right)}{\Gamma(N )},
\end{align}
where \(u = \frac{N(\omega_1 - \sigma^2)}{\alpha_\will P_s \lambda_{\src\eva}}\) and \(v = \frac{N(\omega_1 - \sigma^2)}{P_s \lambda_{\src\eva}}\).
To find the optimal threshold $\omega_1^{\star}$, we differentiate $\dep[1]$ with respect to $\omega_1$ as 
\begin{align}
    \frac{\partial \dep[1] }{\partial \omega_1} &= \frac{\partial}{\partial u} \frac{\Gamma\left(N  , u \right)}{\Gamma(N )} \frac{\partial u}{\partial \omega_1} - \frac{\partial}{\partial v}\frac{\Gamma\left(N  , v \right)}{\Gamma(N )}\frac{\partial v}{\partial \omega_1}\nonumber \\
     &= \frac{N v^{N -1}\exp(-v)}{\Gamma(N )P_s \lambda_{\src\eva}}-\frac{N u^{N -1}\exp(-u)}{\Gamma(N )\alpha_\will P_s \lambda_{\src\eva}}. 
\end{align}
Subsequently, by solving \(\frac{\partial \dep[1] }{\partial \omega_1} = 0\), we get the optimal solution for the optimization problem in \eqref{dep:scheme1-1} as
\begin{align}
\label{ome:close-opt-1}
    \omega_1^{\star} =  \frac{ \alpha_\will\ln(1/\alpha_\will)}{1-\alpha_\will}\lambda_{\src\eva}P_s + \sigma^2.
\end{align}
By substituting \(\omega_1^{\star}\) into \eqref{dep:close-1}, we obtain the minimum DEP as
\begin{align}
\label{dep:close-opt-1}
    \dep^{\star}[1] &= 1 - \frac{1}{\Gamma(N)}\left[ \Gamma\left(N , N\frac{  \alpha_\will\ln(1/\alpha_\will) }{1-\alpha_\will}  \right)\right.\nonumber\\
    &\quad \left.- \Gamma\left( N ,N\frac{ \ln(1/\alpha_\will)}{1-\alpha_\will}   \right)\right].
\end{align}

\begin{remark}
    \normalfont
    Based on \eqref{dep:close-opt-1}, it is evident that the derived minimum DEP is primarily governed by two key parameters: 1) the number of samples \(N\) that Eve utilizes to distinguish the covert signal from her received signal, and 2) the PA coefficient \(\alpha_\will\) assigned by Sam to the public signal \(s_\will\). Since \(N\) is inherently tied to Eve’s monitoring activity, it remains beyond the control of the main system (i.e., Sam). Fortunately, the minimum DEP can be influenced by adjusting \(\alpha_\will\). Therefore, optimizing \(\alpha_\will\) to maximize the minimum DEP at Eve becomes a crucial design objective, as formalized in the following theorem.
\end{remark}

\begin{theorem}\label{theo1}\normalfont
    Let \(\alpha_\will^{\#}\) denote the root of the equation \(\dep^{\star}[1] = \dep^{\rm th}\).
    Then, the effective region for the PA coefficient \(\alpha_\will\) that satisfies the DEP constraint is given by
    \begin{align}
    \label{eq:theo1}
        \alpha_\will^{\#} \leq \alpha_\will \leq 1.
    \end{align}
    \begin{proof}
     See~Appendix~\ref{AppendixA}.
    \end{proof}
\end{theorem}

\begin{remark}
    \normalfont
    Theorem~\ref{theo1} shows that, given a required minimum DEP threshold and assuming that Eve employs an arbitrary number of sample tests, we can analytically determine the appropriate PA coefficient for the public signal to enhance the robust protection of covert information against Eve’s surveillance. Notably, since \(\alpha_\will\) is inherently bounded within a feasible interval, selecting its value poses no practical difficulty; it merely requires choosing an appropriate value from a predefined range.
\end{remark}

\subsection{Eve's DEP in The Second Phase} 
To solve problem \eqref{dep:scheme1-2}, we derive the DEP at Eve as
\begin{align}
\label{dep:scheme1-2-def}
    \dep[2] 
    &= \Pr(\beta_\tom P_r \tau_{\relay\eva} + P_s\tau_{\src\eva} + N\sigma^2 > N\omega_2)\nonumber\\
    &\quad + \Pr(P_r \tau_{\relay\eva} + P_s\tau_{\src\eva} + N\sigma^2 < N\omega_2).
\end{align}
Let \(\chi(x) =  x P_r \tau_{\relay\eva} + P_s\tau_{\src\eva} \). By employing Appendix~\ref{AppendixB} to derive \(F_{\chi(\beta_\tom)}(z)\) in \eqref{CDF-chi-x-close}, the DEP in \eqref{dep:scheme1-2-def} can be obtained as
\begin{figure*}[!t]
    \begin{align}
    \label{CDF-chi-x-close}
    F_{\chi(x)}(z) 
    &= 
   \frac{1}{\Gamma(N)}\left( \frac{xP_r\lambda_{\relay\eva}}{P_s \lambda_{\src\eva}} \right)^{N}
   \calH^{0,1;1,0;1,1}_{1,0;1,1;1,2}            
    \left[\left. \begin{matrix} 1 - 2N;1,1\\-
    \end{matrix}\right|\left. \begin{matrix}    (1,1)\\(0,1)    \end{matrix}\right|\left. \begin{matrix}
    (1-N,1)\\([0,1-2N],1)    
    \end{matrix}\right| \frac{xP_r\lambda_{\relay\eva}}{z} ; \frac{xP_r\lambda_{\relay\eva}-P_s \lambda_{\src\eva}}{P_s \lambda_{\src\eva}} \right].
\end{align}
\hrulefill
\end{figure*}
\begin{align}
\label{dep:close-2}
    \dep[2] = 1-F_{\chi(\beta_\tom)}(N[\omega_2-\sigma^2]) + F_{\chi(1)}(N[\omega_2-\sigma^2]).
\end{align}

Although the expression for the DEP in \eqref{dep:close-2} provides an accurate result, it is challenging to directly solve the optimization problem in \eqref{dep:scheme1-2} using \eqref{dep:close-2} because it is analytically intractable.
To overcome this, we propose to approximate the CDF of the random variable \(\chi(x)\) using the moment matching method similar to \cite{tu2024multihop}. Specifically, we begin with calculating the mean and variance of \(\chi(x)\) as follows:
\begin{align}
    \mathbb{E}\{\chi(x)\} & =  \mathbb{E}\{ x P_r \tau_{\relay\eva} + P_s\tau_{\src\eva}\}=   x P_r \mathbb{E}\{\tau_{\relay\eva}\}+  P_s \mathbb{E}\{\tau_{\src\eva}\}\nonumber\\
    &=   x P_r \sum_{n=1}^N \mathbb{E}\{|h^n_{\relay\eva}|^2\}+  P_s \sum_{n=1}^N\mathbb{E}\{|h^n_{\src\eva}|^2 \}\nonumber\\
    &=   x P_r N \lambda_{\relay\eva} +  P_s N\lambda_{\src\eva}, \\
    \mathbb{V}\{\chi(x)\} & =  \mathbb{V}\{ x P_r \tau_{\relay\eva} + P_s\tau_{\src\eva}\}\nonumber\\
    &= (x P_r )^2 \sum_{n=1}^N \mathbb{V}\{|h^n_{\relay\eva}|^2\} + (P_s)^2\sum_{n=1}^N\mathbb{V}\{|h^n_{\src\eva}|^2 \}  \nonumber\\
    &= N (x P_r \lambda_{\relay\eva})^2 + N (P_s\lambda_{\src\eva})^2.
\end{align}
Subsequently, we fit \(\chi(x)\) into Gamma distribution, whose CDF is given by
\begin{align}
\label{cdf_chi_new}
    F_{\chi(x)}(z) = 1 -\frac{1}{\Gamma(\kappa(x))}\Gamma\left( \kappa(x), \frac{z}{\theta(x)}\right),
\end{align}
where $\kappa(x)$ and $\theta(x)$ are the shape and scale parameters of the approximated gamma distribution, given respectively as
\begin{align}
 \hspace{-3pt}   \kappa(x) & = \frac{\mathbb{E}\{\chi(x)\}^2}{\mathbb{V}\{\chi(x)\}}=\frac{N(P_s  \lambda_{\src\eva} + x P_r  \lambda_{\relay\eva})^2}{ (P_s\lambda_{\src\eva})^2 +  (x P_r \lambda_{\relay\eva})^2} , \\
  \hspace{-3pt}   \theta(x) & = \frac{\mathbb{V}\{\chi(x)\}}{\mathbb{E}\{\chi(x)\}} =\frac{ (P_s\lambda_{\src\eva})^2 +  (x P_r \lambda_{\relay\eva})^2}{P_s  \lambda_{\src\eva} + x P_r  \lambda_{\relay\eva}}.
\end{align}

Using this CDF approximation, \eqref{dep:close-2} becomes differentiable, allowing us to compute its partial derivative with respect to \(\omega_2\), expressed as
\begin{align}
    \frac{\partial \dep[2] }{\partial \omega_2}
    &= \frac{\partial}{\partial \omega_2}\left\{ \frac{1}{\Gamma(\kappa(\beta_\tom))}\Gamma\left( \kappa(\beta_\tom), \frac{N(\omega_2-\sigma^2)}{\theta(\beta_\tom)}\right)\right\}\nonumber  \\
     & \quad  - \frac{\partial}{\partial \omega_2}\left\{ \frac{1}{\Gamma(\kappa(1))}\Gamma\left( \kappa(1), \frac{N(\omega_2-\sigma^2)}{\theta(1)}\right)\right\}\nonumber\\
     &= -\frac{[N(\omega_2-\sigma^2)]^{\kappa(\beta_\tom)-1}}{\Gamma(\kappa(\beta_\tom))\theta(\beta_\tom) } \exp\left(- \frac{\omega_2-\sigma^2}{\theta(\beta_\tom)}\right)\nonumber \\
         & \quad+ \frac{[N(\omega_2-\sigma^2)]^{\kappa(1)-1}}{\Gamma(\kappa(1))\theta(1) }\exp\left(- \frac{\omega_2-\sigma^2}{\theta(1)}\right).
\end{align}
Setting the derivative to zero and solving \(\frac{\partial \dep[2] }{\partial \omega_2} = 0\) leads to the optimal threshold $\omega_2^\star$ as
\begin{align}
\label{ome:close-opt-2}
    &\omega_2^\star 
    =  \frac{\theta(\beta_\tom) \theta(1)[\kappa(1)-\kappa(\beta_\tom)]}{N(\theta(1)-\theta(\beta_\tom))}  \calW\left[ \frac{[ \theta(1)-\theta(\beta_\tom)]/\theta(\beta_\tom)}{[\kappa(1)-\kappa(\beta_\tom)]\theta(1)}\right.\nonumber\\ 
    & \left.\left(\frac{\Gamma(\kappa(1))\theta(1)^{\kappa(1)}}{\Gamma(\kappa(\beta_\tom))\theta(\beta_\tom)^{\kappa(\beta_\tom)}} \right)^{\frac{1}{\kappa(1)-\kappa(\beta_\tom)}} \right] + \sigma^2.
\end{align}
By substituting \(\omega_2^\star[2] \) into \eqref{dep:close-2}, we obtain the minimum DEP for the second communication phase.

\begin{remark}
    \normalfont
    Based on the results in \eqref{ome:close-opt-2} and \eqref{dep:close-2}, it can be observed that the minimum DEP in the second communication phase is not dominated by the noise variance \(\sigma^2\). Instead, it is determined by two key factors: 1) the PA coefficient \(\beta_\tom\) assigned to the public signal \(s_\tom\), and 2) the shape \(\kappa(x)\) and scale \(\theta(x)\) parameters of the Gamma approximation distribution. These parameters encapsulate several critical system parameters, including the total number of sample tests at Eve \((N)\), the transmit powers of Sam \((P_s)\) and Roy \((P_r)\), and the scale fading coefficients of the \({\src\to\eva}\) and \({\relay\to\eva}\) links, denoted by \(\lambda_{\src\eva}\) and \(\lambda_{\relay\eva}\), respectively. Among these parameters, it is evident that adjusting the value of \(\beta_\tom\) not only directly influences \(\omega_2^\star[2]\), but also plays a pivotal role in shaping the Gamma distribution parameters \(\kappa(x)\) and \(\theta(x)\). This dual impact indicates that \(\beta_\tom\) serves as a critical control parameter in determining the minimum DEP. Consequently, it is essential to investigate how to properly define the admissible range of \(\beta_\tom\) to satisfy the minimum DEP constraint while maintaining overall system performance. The following theorem provides further details on this task.
\end{remark}

\begin{theorem}\label{theo2}\normalfont
    Let \(\beta_\tom^{\#}\) denote the root of equation \(\dep^{\star}[2] = \dep^{\rm th}\), the effective region for the PA coefficient \(\beta_\tom\) that satisfies the DEP constraint is given by
    \begin{align}
    \label{eq:theo2}
        \beta_\tom^{\#} \leq \beta_\tom \leq 1.
    \end{align}
    \begin{proof}
    Since \(\beta_\tom > \beta_\bob\) and \(\beta_\tom + \beta_\bob =1\), the range of \(\beta_\tom\) is bounded within the interval \([0.5,1]\). 
    Similar to Theorem~\ref{theo1}, we also derive that $\partial\dep^{\star}[2]/\partial \beta_\tom >0$, for $ \beta_\tom \in [0.5,1]$.
    As a result, $\dep^{\star}[2]$ is a monotonically increasing function with respect to $\beta_\tom \in [0.5,1]$.
    Furthermore, as \(\beta_\tom\to 1\), we can readily observe that \(\chi(\beta_\tom) \to \chi(1)\), which leads to
\begin{align}
    \dep^\star[2] \overset{\beta_\tom\to 1}{=} 1&-F_{\chi(\beta_\tom)}(N[\omega_2^\star-\sigma^2]) \nonumber\\
    &+ F_{\chi(1)}(N[\omega_2^\star-\sigma^2])=1.
\end{align}
Meanwhile, as \(\beta_\tom\to 0.5\), we observe that \(\chi(0.5) < \chi(1)\) and \(F_{\chi(0.5)}(\bullet) > F_{\chi(1)}(\bullet)\), which yields
\begin{align}
\label{dep:close-opt-2}
   \dep^\star[2] \overset{\beta_\tom\to 0.5}{=} 1 &-F_{\chi(\beta_\tom)}(N[\omega_2^\star-\sigma^2])\nonumber\\
   &+ F_{\chi(1)}(N[\omega_2^\star-\sigma^2])<1.
\end{align}
This means that \(\dep^\star[2]\) increases with an increase in \(\beta_\tom\) until achieving the peak value of \(\dep^\star[2] =1\). 
When \(\beta_\tom^{\#}\) is the root of the equation \(\dep^{\star}[2] = \dep^{\rm th}\), the effective operating region of \(\beta_\tom\) becomes \eqref{eq:theo2}, concluding the proof.
    \end{proof}
\end{theorem}

\begin{remark}
    \normalfont
    Theorem~\ref{theo2} reveals that, for any required minimum DEP threshold, an appropriate value of \(\beta_\tom\) can be freely selected from the feasible set, and this selection does not present any practical challenge.
\end{remark}

\section{Secrecy Outage Probability Framework}\label{sec4}
In this section, we evaluate the SOP of Bob's confidential communication considering two eavesdropping scenarios in which the adversary (Eve) employs either SC or MRC to intercept Bob’s information.

\subsection{Preliminary} 
In PLS, the SOP quantifies the probability that a secrecy outage occurs, i.e., the event in which an eavesdropper successfully intercepts and decodes confidential information \cite{TuIOT2024}. 
This probability is closely tied to a predefined secure target rate, which defines the minimum transmission rate at which secrecy must be maintained for legitimate communication.
In the context of this system, the secure target rate of Bob's information is denoted by \(R_\bob\). This rate sets a benchmark that the system must exceed in terms of secrecy capacity to guarantee secure communication between the legitimate sender and receiver. 
The secrecy capacity is defined as
\begin{align}
\label{eq:SEC}
\sec  = \left[\calR_{\bob} - \calR_{\eva}\right]^+, 
\end{align} 
where \(\calR_{\bob}\) is the Bob's capacity over two phases of cooperative transmission and \(\calR_{\eva}\) is the eavesdropper’s capacity.
A secrecy outage occurs when \(\sec < R_\bob\), i.e., the actual secrecy capacity falls short of the target, leading to potential information leakage. In the following, we determine the achievable rates \(\calR_{\bob}\) and \(\calR_{\eva}\).

In the first-phase communication, from \eqref{Roy-SNR-1-1}, the achievable rates for decoding \(s_\will\) and \(c_\bob\) at Roy (with \(s_\will\) eliminated via SIC) are given, respectively, by
\begin{align}
    \calR_{\relay}^{s_\will} = 0.5\log_2\left( 1 + \gamma_{\relay}^{s_\will} \right), \calR_{\relay}^{c_\bob} = 0.5\log_2\left( 1 + \gamma_{\relay}^{c_\bob} \right).
\end{align}
Accordingly, the effective achievable rate for decoding \({c_\bob}\) at Roy in this phase is determined by
\begin{align}
\label{rate-cond-1}
    \calR[1] = 0.5\log_2\left( 1 + \gamma_{\relay} \right), \gamma_{\relay} = \min\{ \gamma_{\relay}^{s_\will},\gamma_{\relay}^{c_\bob}\}.
\end{align}

In the second-phase communication, from \eqref{Bob-SNR-1-1}, the achievable rates for decoding \(s_\tom\) and \(c_\bob\) at Bob (with \(s_\tom\) eliminated via SIC) are expressed, respectively, as
\begin{align}
    \calR_{\bob}^{s_\tom} = 0.5\log_2\left( 1 + \gamma_{\bob}^{s_\tom} \right),\calR_{\bob}^{c_\bob} = 0.5\log_2\left( 1 + \gamma_{\bob}^{c_\bob} \right).
\end{align}
Accordingly, the effective achievable rate for decoding \(c_\bob\) at Bob in this phase is determined by
\begin{align}
\label{rate-cond-2}
    \calR[2]  = 0.5\log_2\left( 1 + \gamma_{\bob} \right), \gamma_{\bob} = \min\{ \gamma_{\bob}^{s_\tom},\gamma_{\bob}^{c_\bob}\}.
\end{align}

From \eqref{rate-cond-1}, \eqref{rate-cond-2}, and \eqref{Eve-SNR}, the end-to-end achievable rates for decoding \(c_\bob\) at Bob and Eve are given, respectively, by
\begin{align}
\label{rate-bob-eve}
    \calR_{\bob} &= 0.5\log_2\left(1+ \min\left\{\gamma_{\relay}, \gamma_{\bob} \right\} \right),\\
    \calR_{\eva} &= 0.5\log_2\left(1 + \gamma_{\eva}\right).
\end{align}

\subsection{SOP Analysis} 
From \eqref{eq:SEC}, the SOP of eavesdropping \(c_\bob\) is measured as
\begin{align}
\label{eq:SOP}
    \sop 
    &=  \Pr\left(\sec \leq R_\bob \right) =  \Pr\left( \calR_{\bob} - \calR_{\eva} \leq R_\bob \right).
\end{align}
This probability quantifies the risk of secrecy failure, which depends on the random channel conditions of both the legitimate and eavesdropping links.

Let \(U = \min\left\{\gamma_{\relay}, \gamma_{\bob} \right\}\), \(V = \gamma_{\eva}\), and \(\overline{\gamma}_{\bob} \triangleq 2^{2R_\bob} - 1\). 
The SOP in \eqref{eq:SOP} can be rewritten as
\begin{align}
\label{eq:SOP-def}
    \sop 
    &=  \Pr\left( \frac{1+ \min\left\{\gamma_{\relay}, \gamma_{\bob} \right\}}{1 + \gamma_{\eva}} \leq 2^{2R_\bob} \right)
    \nonumber\\
    &= \Pr\left( 1+ U \leq 2^{2R_\bob}(1 + V) \right)\nonumber\\
    &= 1 -\Pr\left( V\leq \frac{1+ U}{2^{2R_\bob}}-1, U > \overline{\gamma}_{\bob} \right)\nonumber\\
    &= 1 - \int_{\overline{\gamma}_{\bob}}^\infty F_{V}\left( \frac{ u - \overline{\gamma}_{\bob} }{2^{2R_\bob}} \right)f_U(u)du.
\end{align}
From the result in \eqref{eq:SOP-def}, it is evident that deriving the solution for the SOP critically depends on three challenging factors. First, it requires the CDF of variable \(V\), which is determined by either the SC or MRC scenario. Second, it involves the PDF of variable \(U\), which is determined by the minimum of the two-hop SNR transmissions. Third, the presence of an integral with a non-zero lower limit often complicates the derivation of the  final SOP solution, particularly when the CDF of \(V\) and the PDF of \(U\) contain special functions. In the following, we provide detailed guidelines on how to derive the SOP. 

Specifically, we begin with the following two lemmas for the CDF of variable \(V\).

\begin{lemma}
\normalfont
    The CDF of \(V\) with SC can be derived as
\begin{align}
\label{CDF:V-SC}
        F_{V}^{\rm SC}(v) & = 1 - \exp\left( - \frac{a  v}{\alpha_\bob}\right) -  \frac{\beta_{\bob} a}{bv + \beta_{\bob} a }\exp\left(- \frac{b v}{\beta_{\bob}}\right)\nonumber\\
        &\; + \frac{\beta_{\bob} a}{bv + \beta_{\bob} a}\exp\left(- \frac{b v^2 }{\beta_{\bob} \alpha_\bob}-\frac{a v}{\alpha_\bob}- \frac{b v}{\beta_{\bob}} \right),
\end{align}
where $a  = {\sigma^2}/{(P_s \lambda_{\src\eva})}$ and $b = {\sigma^2}/{(P_r \lambda_{\relay\eva})}$. 
    \begin{proof}
    By invoking \eqref{Eve-SNR-1-1}, \eqref{Eve-SNR-1-2}, and \eqref{eq:SC}, we get
    \begin{align}
        F_{V}(v) 
        &= \Pr\left( V<v\right) = \Pr\left(\max\{\gamma_{\eva}^{c_\bob}[1],\gamma_{\eva}^{c_\bob}[2]\}<v\right) \nonumber\\
        & = \Pr\left(\gamma_{\eva}^{c_\bob}[1] < v,\gamma_{\eva}^{c_\bob}[2]<v\right) \nonumber\\
        & = \Pr\left(|h_{\src\eva}|^2 < \frac{v\sigma^2}{\alpha_\bob P_s },|h_{\relay\eva}|^2 <  \frac{v( P_s|h_{\src\eva}|^2 + \sigma^2)}{\beta_{\bob} P_r }\right)\nonumber\\
        & = \int_0^{\frac{v\sigma^2}{\alpha_\bob P_s }}F_{|h_{\relay\eva}|^2}\left(\frac{v( P_s x + \sigma^2)}{\beta_{\bob} P_r }\right)f_{|h_{\src\eva}|^2}(x)dx\nonumber\\
        & = F_{|h_{\src\eva}|^2}\left( \frac{v\sigma^2}{\alpha_\bob P_s }\right) -  \exp\left(- \frac{\sigma^2 v}{\beta_{\bob} P_r \lambda_{\relay\eva}}\right)\nonumber\\
        &\; \times \int_0^{\frac{v\sigma^2}{\alpha_\bob P_s }}\exp\left(- \frac{v P_s x}{\beta_{\bob} P_r \lambda_{\relay\eva}}\right) \exp\left(- \frac{x}{\lambda_{\src\eva}}\right)\frac{dx}{\lambda_{\src\eva}}.\nonumber
    \end{align}
    After some mathematical manipulations, we get \eqref{CDF:V-SC}. The proof is completed.
    \end{proof}
\end{lemma}

\begin{lemma}
\normalfont
    The PDF of \(V\) with MRC can be derived as
\begin{align}
    \label{CDF:V-MRC}
    &F_{V}^{\rm MRC}(v) =  1 - \exp\left(- \frac{a  v}{\alpha_\bob}\right) - \exp\left(-\frac{bv}{\beta_{\bob}}\right)I(v),
\end{align}
    where 
    \begin{align}
    \label{eq:I(v)}
        I(v)
        &= \sum_{m=0}^\infty \frac{\alpha_{\bob}^m}{m!a}\left(\frac{\beta_\bob }{b}\right)^{m+1}   J(v,m),\\
        J(v,m)
        &=  \frac{\gamma\left( 2m+1 , v\zeta(v)  \right)}{\left(v-\alpha_\bob+\beta_{\bob}a/b \right)^{2m+1}}  ,\\
        \zeta(v) &= {a}/{\alpha_{\bob}}   -  {b }/{\beta_{\bob} } +  {bv}/{(\beta_{\bob}\alpha_{\bob} )}.
    \end{align}

    \begin{proof}
    By invoking \eqref{Eve-SNR-1-1}, \eqref{Eve-SNR-1-2}, and \eqref{eq:MRC}, we get
    \begin{align}
        F_{V}(v) 
        & = \Pr\left(\gamma_{\eva}^{c_\bob}[1] + \gamma_{\eva}^{c_\bob}[2]<v\right) \nonumber\\
        & = \Pr\left(\frac{\alpha_\bob P_s }{\sigma^2}|h_{\src\eva}|^2 + \frac{\beta_{\bob} P_r |h_{\relay\eva}|^2}{ P_s|h_{\src\eva}|^2 + \sigma^2} <  v\right)\nonumber\\
        & = F_{|h_{\src\eva}|^2}\left(\frac{v\sigma^2}{\alpha_\bob P_s}\right) - \int_0^{\frac{v\sigma^2}{\alpha_\bob P_s}}f_{|h_{\src\eva}|^2}(x)\nonumber\\
        &\;\times\left[1-F_{|h_{\relay\eva}|^2}\left(\frac{\left(\sigma^2v-\alpha_\bob P_s x\right)}{\beta_{\bob} P_r/ \left( P_sx/\sigma^2 + 1\right)}\right)\right]dx\nonumber\\
        &= 1 - \exp\left(- \frac{a  v}{\alpha_\bob}\right) - \exp\left(-\frac{bv}{\beta_{\bob}}\right) {\int_0^{\frac{v\sigma^2}{\alpha_{\bob} P_s}}}\frac{1}{\lambda_{\src\eva}} \nonumber\\
        &\;\times\exp\left(-\frac{x}{\lambda_{\src\eva}}-\frac{bP_s(v-\alpha_\bob )}{\sigma^2\beta_{\bob}}x+\frac{b\alpha_\bob P_s^2}{\beta_{\bob}\sigma^4}x^2\right)dx.
\end{align}
To the best of the authors' knowledge, no standard closed-form expression exists for the above integral. To address this challenge, we propose applying the Taylor series expansion of \(f(x) = \exp(x) = \sum_{m=0}^{\infty}\frac{x^m}{m!}\), which enables us to rewrite the integral as
\begin{align}
    I(v) 
        &= \frac{1}{\lambda_{\src\eva}}\sum_{m=0}^\infty \frac{1}{m!}\left(\frac{b\alpha_\bob P_s^2}{\beta_{\bob}\sigma^4}\right)^m\int_0^{\frac{v\sigma^2}{\alpha_{\bob} P_s}} x^{2m}\exp\left( -\frac{x}{\lambda_{\src\eva}} \right)\nonumber\\
        &\;\times \exp\left( -\frac{bP_s(v-\alpha_\bob )}{\sigma^2\beta_{\bob}}x \right)dx,
\end{align}
which has a standard form in \cite[(3.351.1)]{Gradshteyn2014}, yielding its solution as \eqref{eq:I(v)}. The proof is completed.
    \end{proof}
\end{lemma}

It is worth noting that, although the result in \eqref{eq:I(v)} involves an infinite series of polynomial terms, convergence is achieved with as few as five terms.
Therefore, it provides a practical and efficient approach to characterizing the CDF of \(V\).

Subsequently, we derive the PDF of \(U\) in \eqref{eq:SOP-def} based on the relation $f_{U}(u) = {\partial}F_{U}(u)/{\partial u}$.
Accordingly, its exact result can be formulated in the following lemma.
\begin{lemma}\normalfont
    The PDF of \(U\) is derived as in \eqref{PDF:U}, where $c = \sigma^2/[P_s\lambda_{\src\relay}]$ and $d = \sigma^2/[P_r\lambda_{\relay\bob}]$.
    \begin{figure*}    
    \begin{align}
    \label{PDF:U}
        f_U(u) &=   \left[\frac{ \alpha_\will c }{(\alpha_\will - u \alpha_\bob)^2} \exp\left(-\frac{c u}{(\alpha_\will - u \alpha_\bob)} \right){\cal H}(\alpha_\bob - \alpha_\will + u \alpha_\bob) 
        + \frac{c }{\alpha_\bob} \exp\left(-\frac{c u}{\alpha_\bob} \right){\cal H}(\alpha_\will - u \alpha_\bob - \alpha_\bob) \right] \nonumber\\
        &\quad\times\left[\exp \left(-\frac{d u}{\beta_\tom -u\beta_\bob} \right){\cal H}(\beta_\bob - \beta_\tom  + u\beta_\bob) 
        + \exp\left(-\frac{d u}{\beta_\bob} \right){\cal H}(\beta_\tom  - u\beta_\bob - \beta_\bob)\right] \\
        &\quad+
        \left[\frac{d \beta_\tom }{(\beta_\tom -u\beta_\bob)^2} \exp \left(-\frac{d u}{\beta_\tom -u\beta_\bob} \right){\cal H}(\beta_\bob - \beta_\tom  + u\beta_\bob)  
        + \frac{d }{\beta_\bob} \exp\left(-\frac{d u}{\beta_\bob} \right){\cal H}(\beta_\tom  - u\beta_\bob - \beta_\bob)  
         \right] \nonumber\\
        &\quad\times\left[\exp\left(-\frac{c u}{(\alpha_\will - u \alpha_\bob)} \right) {\cal H}(\alpha_\bob - \alpha_\will + u \alpha_\bob)
        + \exp\left(-\frac{c u}{\alpha_\bob} \right){\cal H}(\alpha_\will - u \alpha_\bob - \alpha_\bob)\right] \nonumber.
    \end{align}
        \hrulefill
    \end{figure*}
    \begin{proof}
        We begin this proof by first recalling \eqref{Roy-SNR-1-1}, \eqref{Bob-SNR-1-1}, \eqref{rate-cond-1}, and \eqref{rate-cond-2} to derive the CDF of \(U\) as
        \begin{align}
        \label{cdf:U}
            F_U(u) &= \Pr\left(U \leq u \right) = 1 - \Pr\left(\min\left\{\gamma_{\relay}, \gamma_{\bob} \right\} \geq u \right) \nonumber\\
            &= 1 - \Pr\left(\gamma_{\relay} \geq  u , \gamma_{\bob} \geq u \right)\nonumber\\
            &= 1 - \Pr\left(\min\{ \gamma_{\relay}^{s_\will},\gamma_{\relay}^{c_\bob}\} >  u , \min\{ \gamma_{\bob}^{s_\tom},\gamma_{\bob}^{c_\bob}\} > u \right)\nonumber\\
            &= 1 - \Pr\left(\gamma_{\relay}^{s_\will}>u,\gamma_{\relay}^{c_\bob}>  u , \gamma_{\bob}^{s_\tom} > u,\gamma_{\bob}^{c_\bob} > u \right)\nonumber\\
            &= 1 - \Pr\left( |h_{\src\relay}|^2 > \frac{u \sigma^2/P_s}{\min\{\alpha_\will-u\alpha_\bob, \alpha_\bob\}} \right) \nonumber\\
            &\qquad\times \Pr\left( |h_{\relay\bob}|^2 > \frac{u \sigma^2/P_r}{\min\{\beta_\tom -u\beta_\bob, \beta_\bob\}} \right) \nonumber\\
            &= 1 - \left[1 - F_{|h_{\src\relay}|^2}\left(\frac{u \sigma^2/P_s }{\min\{\alpha_\will-u\alpha_\bob, \alpha_\bob\}} \right)\right]\nonumber\\
            &\qquad\times\left[1 - F_{|h_{\relay\bob}|^2}\left(\frac{u \sigma^2/P_r}{\min\{\beta_\tom -u\beta_\bob, \beta_\bob\}} \right) \right], 
        \end{align}
        where \(u < \min\{ {\alpha_\will}/{\alpha_\bob}, {\beta_\tom}/{\beta_\bob} \}\); otherwise, \( F_U(u) =1\).

        To remove the discontinuity introduced by the piecewise min functions in \eqref{cdf:U}, the Heaviside step function \({\cal H}(\bullet)\) is employed as
        \begin{align}
        F_U(u) = 1 &- \left[\exp\left(-\frac{c u}{(\alpha_\will - u \alpha_\bob)} \right) {\cal H}(\alpha_\bob - \alpha_\will + u \alpha_\bob)\right.\nonumber\\ 
        &\left. + \exp\left(-\frac{c u}{\alpha_\bob} \right){\cal H}(\alpha_\will - u \alpha_\bob - \alpha_\bob)\right]\nonumber\\
        &\times\left[\exp \left(- \frac{d u}{\beta_\tom -u\beta_\bob} \right){\cal H}(\beta_\bob - \beta_\tom  + u\beta_\bob)\right.\nonumber\\ 
        &\left. + \exp\left( - \frac{d u}{\beta_\bob} \right){\cal H}(\beta_\tom  - u\beta_\bob - \beta_\bob)\right].
    \end{align}
    By taking the derivative for \(F_U(u)\) with respect to \(u\), we get \eqref{PDF:U}. The proof is concluded.
    \end{proof}
\end{lemma}

\begin{theorem}
    \label{theorem3}
    The SOP of eavesdropping \(c_\bob\) can be given by
    \begin{align}\label{eq:SOP_cB}
    \sop 
    \approx 1 &- \frac{(\Theta-\overline{\gamma}_{\bob})\pi}{2M}\sum\limits_{m=1}^M \sqrt{1 - \delta_m^2} f_U\left( (\delta_m + 1){\Theta}/{2} \right)\nonumber\\
    &\times F_{V}\left( \frac{\Theta-\overline{\gamma}_{\bob}}{ 2^{2R_\bob + 1}}(\delta_m+1)  \right),
\end{align}
where \(\Theta \triangleq \min ( \alpha_\will/\alpha_\bob,\beta_\tom/\beta_\bob ) \), \( \delta_m = \cos\left( {(2m - 1)\pi}/{[2M]} \right) \), and \(M\) is the trade-off parameter.
    \begin{proof}
   Since \(F_U(u)\) in \eqref{cdf:U} has the feasible domain \(0 < u < \Theta\) and the integral in \eqref{eq:SOP-def} is only valid on the feasible domain \(\overline{\gamma}_{\bob} < u < \infty\), we can rewrite the SOP in \eqref{eq:SOP-def} as
\begin{align}\label{eq:temp56}
    \sop
    = 1 &- \int_{\overline{\gamma}_{\bob}}^{\Theta} F_{V}\left( \frac{ u - \overline{\gamma}_{\bob} }{2^{2R_\bob}} \right)f_U(u)du.
\end{align}
Due to the complicated forms of \(F_{V}\left(\bullet \right)\) and \(f_{U}\left(\bullet \right)\), it is challenging to derive the exact closed-form expression for \eqref{eq:temp56}. Therefore, we approximate \eqref{eq:temp56} by using the Gaussian--Chebyshev quadrature method \cite{vu2021cooperative,tu2023short}, which yields \eqref{eq:SOP_cB}. We conclude the proof.
    \end{proof}
\end{theorem}

\section{Covert Rate Maximization Framework}\label{sec5}
In this section, we jointly optimize the PA coefficients \(\boldsymbol{\alpha}= \{\alpha_\will,\alpha_\bob\}\) and \(\boldsymbol{\beta}=\{\beta_\tom,\beta_\bob\}\) to maximize the covert communication rate, while simultaneously ensuring the QoS requirements of legitimate users, maintaining covert transmission against eavesdropping detection, and supporting SIC-based NOMA procedures.
\subsection{Problem Formulation}
The optimization problem is mathematically formulated as
\begin{subequations}
\label{problem-2}
    \begin{align}
\label{problem-2a}
       &\max_{\boldsymbol{\alpha},\boldsymbol{\beta}}
        \; \min\{\calR_{\relay}^{c_\bob},\calR_{\bob}^{c_\bob}\}\\
\label{problem-2b}
       \text{s.t.} \;& \min\{ \calR_{\relay}^{s_\will},\calR_{\will}^{s_\will}\}  \ge r_\will ,\\
\label{problem-2c}
       &\min\{ \calR_{\bob}^{s_\tom},\calR_{\tom}^{s_\tom}\}\ge r_{\tom},\\
\label{problem-2d}
         & \alpha_\will + \alpha_\bob \leq 1,\quad \beta_\tom + \beta_\bob \leq 1,\\
\label{problem-2e}
         & \alpha_\will^{\#} \leq \alpha_\will \leq  1,\quad \beta_\tom^{\#} \leq \beta_\tom \leq  1,
    \end{align}
\end{subequations}
where \(r_\will\) and \(r_\tom\) denote the minimum rate requirements when decoding the public signals $s_\will$ and $s_\tom$, respectively, whereas $\calR_{\will}^{s_\will}$ and $\calR_{\tom}^{s_\tom}$ are determined based on \eqref{Will-SNR-1-1} and \eqref{Tom-SNR-1-1} as
\begin{align}
    \calR_{\will}^{s_\will}=0.5\log_2\left( 1 + \gamma_{\will}^{s_\will} \right), \calR_{\tom}^{s_\tom} =0.5\log_2\left( 1 + \gamma_{\tom}^{s_\tom} \right).
\end{align}
In \eqref{problem-2}, the constraints \eqref{problem-2b} and \eqref{problem-2c} ensure the minimum QoS requirements for Willie and Tom, respectively.
The constraints \eqref{problem-2d} enforce feasible PA budgets, while the constraints \eqref{problem-2e} guarantee the effective regions of the PA coefficients in which Eve’s DEP remains above the required threshold \(\dep^{\rm th}\). 
It is worth mentioning that the transmission rate between \(\src\) and \(\will\) in the second phase (i.e., transmission of \(\widehat{s}_\will\)) with the outage rate target \(\widehat{r}_\will\) is excluded from \eqref{problem-2} since it can be readily handled by solving
\begin{align}
\label{rate-phase-2}
    0.5\log_2\left( 1 + \gamma_\will^{\widehat{s}_\will} \right) \ge \widehat{r}_\will \Leftrightarrow  P_r\leq \frac{\left(\frac{P_s |h_{\src\will}|^2}{(2^{2\widehat{r}_\will}-1)} - \sigma^2\right)}{|h_{\relay\will}|^2 }.
\end{align}

To solve the optimization problem \eqref{problem-2}, we first rewrite the constraints in \eqref{problem-2b} and \eqref{problem-2c} as
\begin{align}
\label{condition-1}
   \eqref{problem-2b}  & \Leftrightarrow \alpha_{\will}  \ge  {\overline{\gamma}_{\will}(P_s  + \sigma^2/l_1)}{/[(1 + \overline{\gamma}_{\will})P_s]} ,\\
\label{condition-2}
  \eqref{problem-2c} & \Leftrightarrow 
     \beta_{\tom}  \ge {\overline{\gamma}_{\tom} (P_r  + \sigma^2/l_2)}{/[(1 + \overline{\gamma}_{\tom})P_r]},
\end{align}
where \(\overline{\gamma}_{\will} \triangleq 2^{2r_{\will}} - 1\), \(\overline{\gamma}_{\tom} \triangleq 2^{2r_{\tom}} - 1\), \(l_1\triangleq\min\{|h_{\src\relay}|^2,|h_{\src\will}|^2\}\), and \(l_2\triangleq\min\{|h_{\relay\tom}|^2,|h_{\relay\bob}|^2\}\).

Notably, the objective function \eqref{problem-2a} is referred to as a fairness problem, which is non-convex. This problem becomes fair if and only if \(\gamma_{\relay}^{c_\bob} = \gamma_{\bob}^{c_\bob} \). By solving this equality in conjunction with \eqref{problem-2d}, i.e., \(\alpha_\bob = 1 -\alpha_\will \) and \(\beta_\bob = 1 -\beta_\tom \), we get
\begin{align}
\label{condition-3}
  \gamma_{\relay}^{c_\bob} = \gamma_{\bob}^{c_\bob} 
  &\Leftrightarrow \alpha_\will = 1 - (1 -\beta_\tom)\frac{ P_r |h_{\relay\eva}|^2}{P_s  |h_{\src\eva}|^2}.
\end{align}

Now, by combining \eqref{condition-1}--\eqref{condition-3} with \eqref{problem-2e}, one can derive the solution for \eqref{problem-2} as
\begin{align}
      1& \geq \beta_{\tom}^\star \geq \max\left\{\beta_\tom^{\#}, \frac{\overline{\gamma}_{\tom}(P_r  + \sigma^2/l_2)}{(1 + \overline{\gamma}_{\tom})P_r}, 1 - \frac{P_s  |h_{\src\eva}|^2}{ P_r |h_{\relay\eva}|^2}\nonumber\right.\\
      &\quad\left.\left( 1 - \max\left\{\alpha_\will^{\#}, \frac{\overline{\gamma}_{\will}(P_s  + \sigma^2/l_1)}{(1 + \overline{\gamma}_{\will})P_s}\right\}\right)\right\},\\      
      \alpha_\will^\star & = 1 - (1 -\beta_\tom^\star)\frac{ P_r |h_{\relay\eva}|^2}{P_s  |h_{\src\eva}|^2}.
      \label{alpha-opt-solution}
\end{align}
We note that \( \gamma_{\relay}^{c_\bob} \) and \(\gamma_{\bob}^{c_\bob}\) are linear functions of \(\alpha_\bob \) and \(\beta_\bob\), respectively; thus, maximizing the objective function \eqref{problem-2a} is equivalent to maximizing  \( \{\alpha_\bob,\beta_\bob\}\). That is to say that \(\{\alpha_\will,\beta_\tom\}\) should be minimized. Therefore, the value of \(\beta_{\tom}^\star\) can be obtained as 
\begin{align}
\label{beta-opt-solution}
       \beta_{\tom}^\star &= \max\left\{\beta_\tom^{\#}, \frac{\overline{\gamma}_{\tom}(P_r  + \sigma^2/l_2)}{(1 + \overline{\gamma}_{\tom})P_r}, 1 - \frac{P_s  |h_{\src\eva}|^2}{ P_r |h_{\relay\eva}|^2}\nonumber\right.\\
      &\quad\left.\left( 1 - \max\left\{\alpha_\will^{\#}, \frac{\overline{\gamma}_{\will}(P_s  + \sigma^2/l_1)}{(1 + \overline{\gamma}_{\will})P_s}\right\}\right)\right\}.
\end{align}

\section{Secrecy Rate Maximization Framework}\label{sec6}
In this section, we formulate the secrecy rate maximization problem for the considered system in the presence of eavesdroppers, while ensuring the QoS requirements of legitimate users and supporting SIC-based NOMA integrity. We then transform the problem into a convex approximation. Finally, an iterative optimization algorithm is proposed to obtain a suboptimal solution.

\subsection{Problem Formulation}
Building upon the secrecy capacity in \eqref{eq:SEC}, we aim to maximize the secrecy rate by jointly optimizing the variables \(\boldsymbol{\alpha}= \{\alpha_\will,\alpha_\bob\}\) and \(\boldsymbol{\beta}=\{\beta_\tom,\beta_\bob\}\), formulated as
\begin{subequations}
\label{problem-1}
    \begin{align}
\label{problem-1a}
       &\max_{\boldsymbol{\alpha},\boldsymbol{\beta}}
        \quad \sec \\
\label{problem-1b}
       \text{s.t.} \;& \min\{ \calR_{\relay}^{s_\will},\calR_{\will}^{s_\will}\}  \ge r_\will ,\\
\label{problem-1c}
       &\min\{ \calR_{\bob}^{s_\tom},\calR_{\tom}^{s_\tom}\}\ge r_{\tom},\\
\label{problem-1d}
         & \alpha_\will + \alpha_\bob \leq 1 , \beta_\tom + \beta_\bob \leq 1.
    \end{align}
\end{subequations}

In problem \eqref{problem-1}, it is evident that the objective function \eqref{problem-1a} and the constraints \eqref{problem-1b} and \eqref{problem-1c} are non-convex due to the non-linearity introduced by the logarithmic functions in the rate expressions.
This non-convexity renders the problem challenging and prevents a direct solution. Specifically, the secrecy rate is formulated as the difference between two logarithmic terms representing the legitimate and eavesdropper channels, which typically results in the difference of concave functions. Such a formulation is generally non-convex and thus difficult to optimize. Moreover, although the logarithmic function itself is concave, its composition with fractional SINR expressions, already non-convex, yields a function that is neither convex nor concave, a phenomenon referred to as logarithmic composition. In addition, the presence of rate-based constraints further introduces non-convexity into the feasible region \cite{vu2025study,Vu2024Feb}, thereby adding to the complexity of the optimization process.
To address this issue, we propose the following alternative approach.

\subsection{Proposed Solution}

\subsubsection{Convexifying the Objective Function}
For handling \eqref{problem-1a}, we first introduce an auxiliary variable \(\phi\) to bound the objective function such that
\begin{align}\label{eq:phi-o}
    \phi \leq \sec. 
\end{align}
Notably, $\sec$ falls within the form of \(\psi(x,y)=\ln(y)-\ln(x)\), which can be rewritten in a compact form of \(\psi(x,y,z) = \ln(y) + \ln(z) + 1 - zx\), where \(z=1/x\). Accordingly, we further introduce four auxiliary variables \(\{ t_{\bob}, t_{\eva}, q, p \}\) and rewrite \eqref{eq:phi-o} as
\begin{subequations}
    \begin{empheq}[left={ \empheqlbrace\,}]{align}
       \phi &\leq \frac{1}{2\ln(2)}\left[\ln\left(1 + t_{\bob}\right) + \ln(q) - p + 1\right],
        \label{eq:phi} \\
       t_{\bob}  &\leq \min\{\gamma_{\relay}^{c_\bob},\gamma_{\bob}^{c_\bob}\},
        \label{eq:t-bob}\\
       t_{\eva} &\leq \gamma_{\eva}^{c_\bob},
        \label{eq:t-eve}\\
       p &\ge q(1 + t_{\eva}).
       \label{eq:t-q}
    \end{empheq}
\end{subequations}

For handling \eqref{eq:phi}, we introduce the following lemma.

\begin{lemma}\normalfont\label{lema1}
    Given \(\{x,\widehat{x}\}>0\) and the functions \(f(x) = \ln(1 + x) \) and \(g(x) = \ln(x) \), the following inequalities hold
    \begin{align}\label{eq:temp69}
        f(x) &= \ln(1 + x) \ge \ln(1 + \widehat{x}) + \frac{\widehat{x}^2}{\widehat{x}+1}\left(\frac{1}{\widehat{x}} - \frac{1}{x}\right),\\
        \label{eq:temp70}
        g(x) &= \ln(x)\ge \ln(\widehat{x}) +  \widehat{x}\left( {1}/{\widehat{x}} - {1}/{x}\right),
    \end{align}
    where \(\widehat{x}\) is the local point of \(x\).
    \begin{proof}        
        By applying the first-order Taylor series expansion for functions \(f(y) = \ln(1 + 1/y) \) and \(g(y) = \ln(1/y) \) at a strictly positive local point \(\widehat{y}\) and mapping \(x = 1/y\) and \(\widehat{x} = 1/\widehat{y}\), we immediately get \eqref{eq:temp69} and \eqref{eq:temp70}. The proof is concluded.
    \end{proof}
\end{lemma}

From \textbf{Lemma}~\ref{lema1}, the constraint \eqref{eq:phi} can be convexified as
\begin{align}
\label{eq:phi_new}
     \phi 
     &\leq  \ln(1 + \widehat{t}_{\bob}) + \frac{\widehat{t}_{\bob}^2}{\widehat{t}_{\bob}+1}\left( {1}/{\widehat{t}_{\bob}} - {1}/{t_{\bob}}\right) - p  + 1\nonumber\\
     &\quad + \ln(\widehat{q}) +  \widehat{q}\left( {1}/{\widehat{q}} - {1}/{q}\right),
\end{align}
where $\widehat{t}_{\bob}$ and $\widehat{q}$ are the feasible points of $t_{\bob}$ and $q$, respectively.

For handling \eqref{eq:t-bob}, we transform the SNR expressions in \eqref{Roy-SNR-1-1} and \eqref{Bob-SNR-1-1} into convex forms as
\begin{subequations}
  \label{eq:t-bob-new}
    \begin{empheq}[left={ \eqref{eq:t-bob} \Leftrightarrow \empheqlbrace\,}]{align}
      & t_{\bob} \leq \alpha_\bob P_s |h_{\src\relay}|^2/\sigma^2,
        \label{eq:t-bob-a} \\
      & t_{\bob}  \leq \beta_\bob P_r |h_{\relay\bob}|^2/\sigma^2.
        \label{eq:t-bob-b}
    \end{empheq}
\end{subequations}

For handling \eqref{eq:t-eve}, under the SC technique, we transform the SNR expression in \eqref{eq:SC} into a convex form as
\begin{subequations}
  \label{eq:t-SC-eve}
    \begin{empheq}[left={\eqref{eq:t-eve} \Leftrightarrow \empheqlbrace\,}]{align}
      & t_{\eva} \leq {\alpha_\bob P_s |h_{\src\eva}|^2}/{\sigma^2}, 
        \label{eq:t-SC-eve-a} \\
      & t_{\eva}(P_s  |h_{\src\eva}|^2 + \sigma^2) \leq \beta_{\bob} P_r |h_{\relay\eva}|^2.
        \label{eq:t-SC-eve-b}
    \end{empheq}
\end{subequations}
Meanwhile, under the MRC technique, we transform the SNR expression in \eqref{eq:MRC} into a convex form as
\begin{align}
  \label{eq:t-MRC-eve}
   \eqref{eq:t-eve} \Leftrightarrow t_{\eva} \leq {\alpha_\bob P_s |h_{\src\eva}|^2}/{\sigma^2} + \beta_{\bob} \frac{P_r |h_{\relay\eva}|^2}{P_s|h_{\src\eva}|^2 + \sigma^2}.
\end{align}

For handling \eqref{eq:t-q}, we introduce the following lemma.

\begin{lemma}\normalfont\label{lema2}
    Given \(\{x,y,z,\widehat{x},\widehat{y}\}>0\) and the function \(f(x,y) = xy\), the following inequality holds
    \begin{align}
    \label{eq:lemma3}
        f(x,y) \leq z \Leftrightarrow  z &\ge \frac{1}{2}\frac{\widehat{y}}{\widehat{x}}x^2 + \frac{1}{2}\frac{\widehat{x}}{\widehat{y}}y^2.
    \end{align}
    where $\widehat{x}$ and $\widehat{y}$ are the local points of $x$ and $y$, respectively.
    \begin{proof}
        Starting from the inequality \((X - Y)^2\ge 0 \Leftrightarrow XY \leq \frac{1}{2}(X^2 + Y^2) \), with \(\{X,Y\}>0\),
        we let \(X = x\widehat{y}\) and \(Y = y\widehat{x}\).
        Dividing both sides of the inequality by \(\widehat{x}\widehat{y}\) yields \(xy \leq \frac{1}{2}\frac{\widehat{y}}{\widehat{x}}x^2 + \frac{1}{2}\frac{\widehat{x}}{\widehat{y}}y^2\), where the equality holds when $x$ and $y$ converge to 
        $\widehat{x}$ and $\widehat{y}$, respectively. 
    \end{proof}
\end{lemma}

From \textbf{Lemma}~\ref{lema2}, the constraint \eqref{eq:t-q} can be convexified as 
\begin{align}
\label{eq:t-q-new}
    p \ge \frac{1}{2}\frac{(1 + \widehat{t}_{\eva})}{\widehat{q}}q^2 + \frac{1}{2}\frac{\widehat{q}}{(1 + \widehat{t}_{\eva})}(1 + t_{\eva})^2.
\end{align}

\subsubsection{Convexifying the Constraints \eqref{problem-1b} and \eqref{problem-1c}}
Similar to Section~\ref{sec5}, the constraints \eqref{problem-1b} and \eqref{problem-1c} can be reformulated as the convex forms of \eqref{condition-1} and \eqref{condition-2}, respectively.

\subsubsection{Iterative Optimization Algorithm}Based on the above analysis, we are now ready to solve the following equivalent convex problems to obtain solutions to the original problem. 

For the scenario in which the SC technique is adopted at Eve, the equivalent optimization problem is expressed as
\begin{align}
\label{problem1-SC}
&\max_{\substack{\boldsymbol{\alpha},\boldsymbol{\beta},\phi}} \; \phi \; \text{s.t.}\; \left\{\begin{matrix}
   \eqref{condition-1}, \eqref{condition-2}, \eqref{problem-1d}, \eqref{eq:phi_new}, \eqref{eq:t-bob-a},\\ \eqref{eq:t-bob-b}, \eqref{eq:t-SC-eve-a}, \eqref{eq:t-SC-eve-b}, \eqref{eq:t-q-new}  
\end{matrix}\right\}.
\end{align}

For the scenario in which the MRC technique is adopted at Eve, the equivalent optimization problem is expressed as
\begin{align}
\label{problem1-MRC}
&\max_{\substack{\boldsymbol{\alpha},\boldsymbol{\beta},\phi}} \; \phi \; \text{s.t.}\; \left\{\begin{matrix}
   \eqref{condition-1}, \eqref{condition-2}, \eqref{problem-1d}, \eqref{eq:phi_new},\\ \eqref{eq:t-bob-a},\eqref{eq:t-bob-b},\eqref{eq:t-MRC-eve},
     \eqref{eq:t-q-new}
\end{matrix}\right\}.
\end{align}
 
The convex approximation problems \eqref{problem1-SC} and \eqref{problem1-MRC} can be effectively solved using interior-point methods or a CVX solver \cite{Vu2024Feb}.
Notably, given the optimal values of \(\{\boldsymbol{\alpha},\boldsymbol{\beta}\}\) obtained from the CVX solver within a single iteration, the objective function may not necessarily converge to its optimal value.
Therefore, we iteratively solve \eqref{problem1-SC} (or \eqref{problem1-MRC}) to generate locally optimal points $\{{\boldsymbol{\alpha}},{\boldsymbol{\beta}},\phi \}$ until the objective function converges, as illustrated in Algorithm~\ref{Algorith1}.
Since \eqref{problem1-SC} and \eqref{problem1-MRC} are maximization problems bounded above by the constraints in \eqref{problem-1d} and \eqref{eq:phi_new}, the convergence of Algorithm~\ref{Algorith1} is guaranteed and the algorithm terminated after a finite number of iterations \cite{tu2024hybrid,tu2025semi}.

\begin{algorithm}[!t]
 \caption{{Adaptive Resource Allocation For Secrecy Achievable Rate Maximization}\label{Algorith1}}

\KwIn{Initialize iteration $k=0$, threshold $\varepsilon= 10^{-2}$, and feasible points 
$(\boldsymbol{\alpha}^{(0)}, \boldsymbol{\beta}^{(0)}, \phi^{(0)})$.}

\KwRep

{1. Set $k=k+1$.}

{2. Solve the problem \eqref{problem1-SC} (or \eqref{problem1-MRC}) by CVX to update $\phi^{(k)}$, $\boldsymbol{\alpha}^{(k)}$, and $\boldsymbol{\beta}^{(k)}$.}

\KwUnt{$|\phi^{(k)} - \phi^{(k-1)} |/\phi^{(k-1)} \leq \varepsilon$.}

\KwOut{Optimal solution  \(\{\boldsymbol{\alpha}^\star,\boldsymbol{\beta}^\star\}\).}
\end{algorithm}

\section{Numerical Results and Discussion}\label{sec7}
This section presents Monte Carlo simulations to validate the accuracy of the developed analytical and optimization frameworks. 
Unless otherwise specified, the simulation parameters are as given in Table~\ref{table:simu_param}.

\begin{table}[!t]
\renewcommand{\arraystretch}{1.25}
\centering
\caption{Summary of simulations parameters.}
\begin{tabular}{|l|c|}
\hline
\multicolumn{1}{|c|}{\centering{\textbf{Parameter}}}  & \multicolumn{1}{c|}{\centering{\textbf{Value}}}\\
\hline
\hline
{Transmit power at Sam} &
{$P_s \in [-10,30]$ dBm}
\\ \hline
{Transmit power at Roy} &
{$P_r$ employed as \eqref{rate-phase-2}}
\\ \hline
{Average channel gains} &
\makecell{$\lambda_{\rm{SR}}=\lambda_{\rm{RB}} =1$, \\$ \lambda_{\rm{SW}} = \lambda_{\rm{SE}} = \lambda_{\rm{RT}} = \lambda_{\rm{RE}} = 0.5$}
\\ \hline
{Normalized noise variance} &
{$\sigma^2=1$}
\\ \hline
{Collected samples at Eve} &
{$N \in [1,30]$}
\\ \hline
{Judgment thresholds} &
{$\{\omega_1,\omega_2\} \in [0,15]$ dB}
\\ \hline
{Secrecy rate threshold} &
{$R_{\rm{B}} \in [0.1,0.4]$ bps/Hz}
\\ \hline
{Target data rate thresholds} &
{$\{r_{\rm{T}},r_{\rm{W}},{\hat r}_{\rm{W}}\} \in [0.1,2.8]$ bps/Hz}
\\ \hline

\hline

\hline
\end{tabular}
\label{table:simu_param}
\end{table}

\subsection{DEP Evaluation}
\begin{figure}[!t]
     \centering
     \includegraphics[width=.85\linewidth]{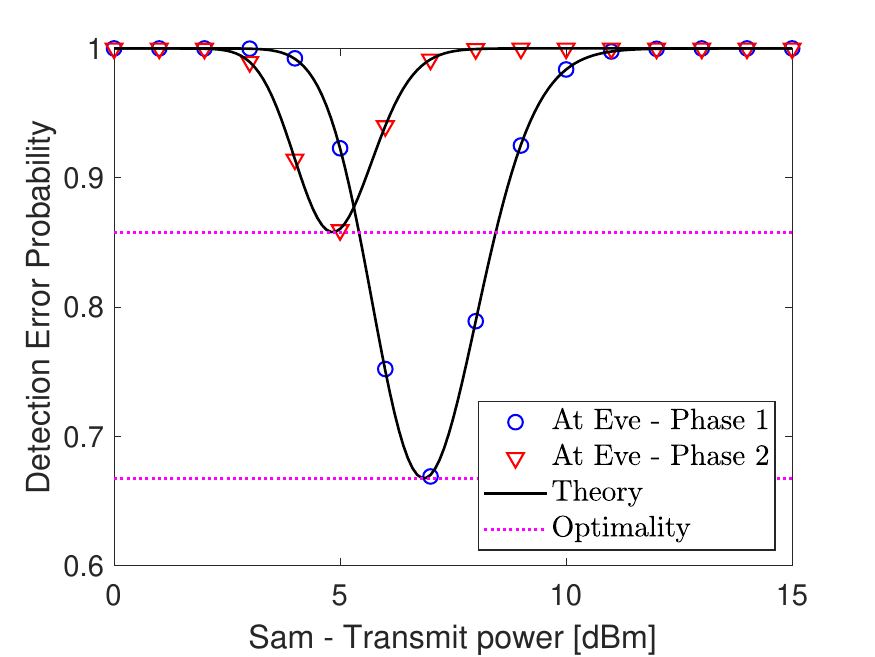}
    \par \footnotesize{a) DEP vs. \(P_s\) at \(N = 15\) and \(\omega_1 = \omega_2 = 5\) dB.}\par
    \includegraphics[width=.85\linewidth]{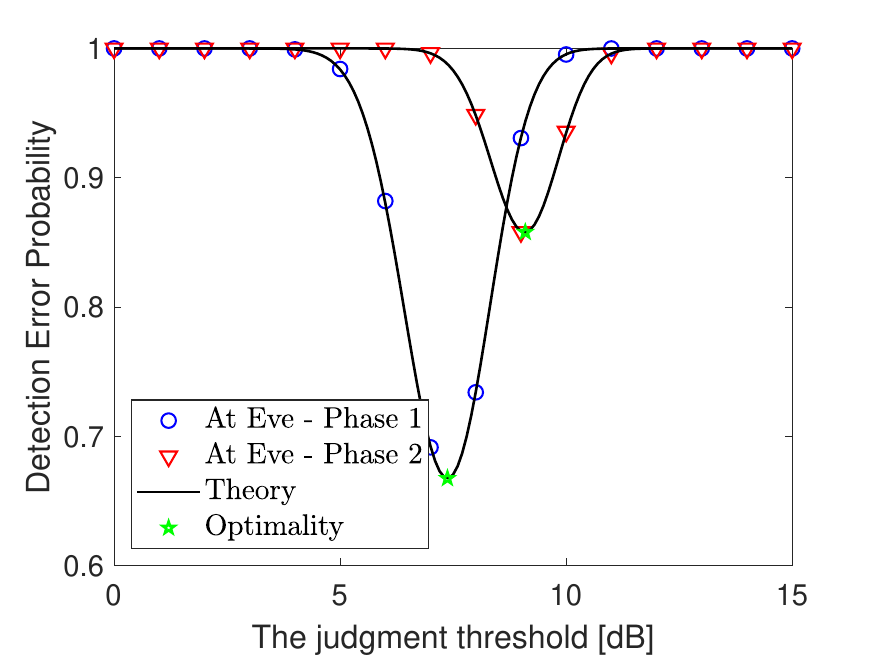}
    \par \footnotesize{b) DEP vs. \(\omega_1=\omega_2\) at \(N = 15\) and \(P_s=10\) dBm.}
    \includegraphics[width=.85\linewidth]{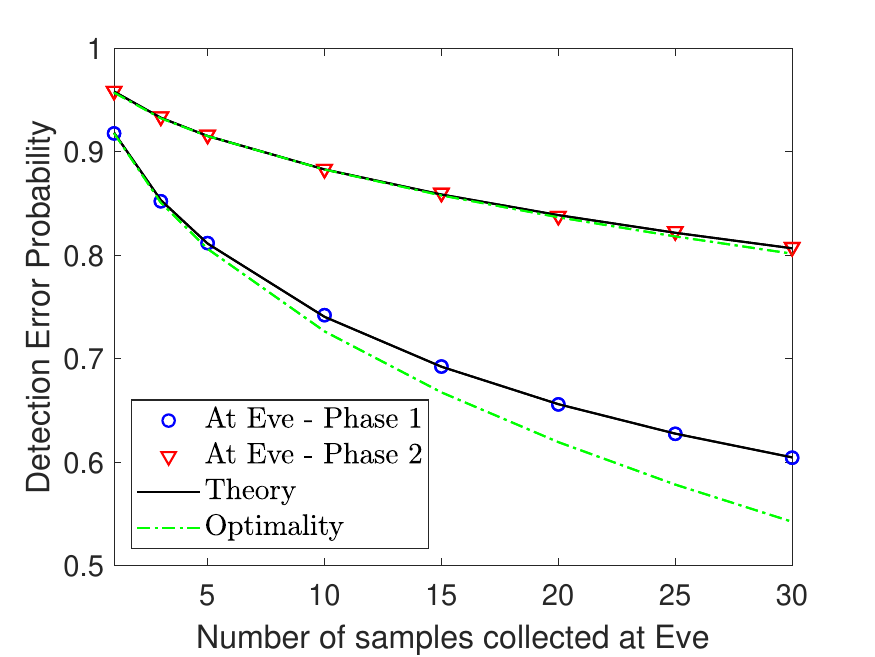}
    \par \footnotesize{c) DEP vs. \(N\) with \(\omega_1 = 7 \) dB, \( \omega_2 = 9 \) dB, and \(P_s=10\) dBm.}
     \caption{DEP performance at Eve with \(\alpha_\bob = \beta_\bob = 0.2 \) and \(P_r = 0.5P_s\).
\label{fig:2}}
\vspace{-0.25cm}
\end{figure}

We first evaluate the DEP of the considered system from the perspective of the warden, i.e., Eve.
Fig.~\ref{fig:2}(a) illustrates the DEP across two phases of covert monitoring. 
It can be observed that when the judgment thresholds are arbitrarily set to \(\omega_1 = \omega_2 = 5\) dB, the DEPs in both phases exhibit a convex downward trend. Specifically, the DEPs initially decrease with increasing values of $P_s$, and subsequently increase as $P_s$ continues to rise.
Notably, the analytical results derived in \eqref{dep:close-1} and \eqref{dep:close-2} closely match the simulation outcomes, which validates the correctness of our analysis.
Furthermore, it is worth mentioning that when Eve successfully optimizes the judgment thresholds $\omega_1$ and $\omega_2$ to minimize the DEP, as described in \eqref{dep:close-opt-1} and \eqref{dep:close-opt-2}, the DEP becomes invariant with respect to $P_s$, as shown by the constant magenta dotted lines in Fig.~\ref{fig:2}(a).
To further demonstrate this, Fig.~\ref{fig:2}(b) depicts the DEP as a function of \(\omega_1 = \omega_2\). It is evident that the closed-form solutions developed in \eqref{ome:close-opt-1} and \eqref{ome:close-opt-2} accurately estimate the minimum DEPs in both phases at Eve. These minimum DEP values also align with the constant lines in Fig.~\ref{fig:2}(a).
Finally, we examine the effect of monitoring sample collection in Fig.~\ref{fig:2}(c).
The results indicate that as Eve increases the number of sample tests for covert signal classification, the DEP decreases.
This improvement is attributed to the enhanced received signal strength resulting from more extensive sampling, which in turn improves Eve’s classification capability.

\subsection{SOP Evaluation}

\begin{figure}[!t]
     \centering
     \includegraphics[width=.85\linewidth]{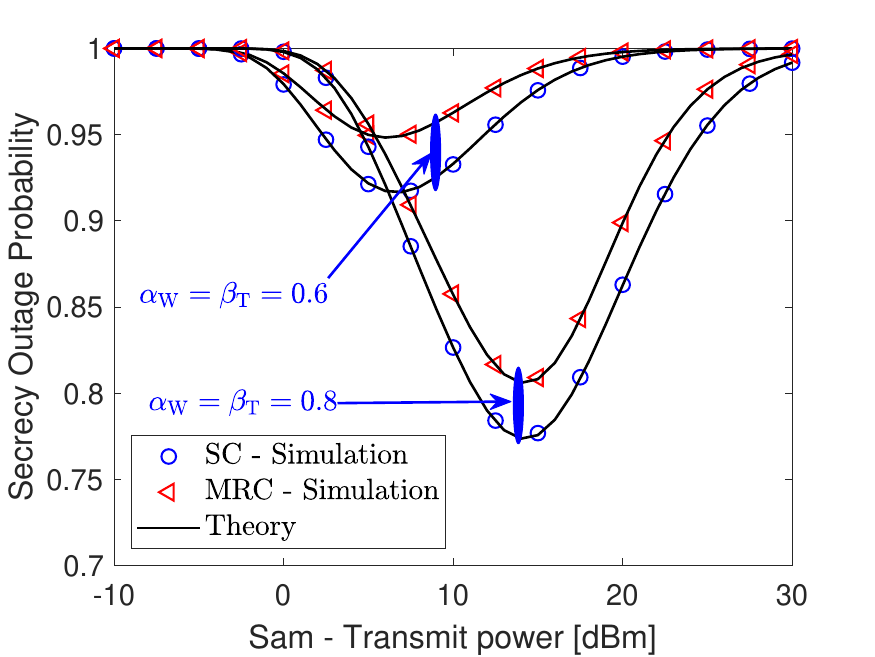}
    \par \footnotesize{a) SOP vs. \(P_s\) at \(R_\bob = 0.2\) bps/Hz.}\par
    \includegraphics[width=.85\linewidth]{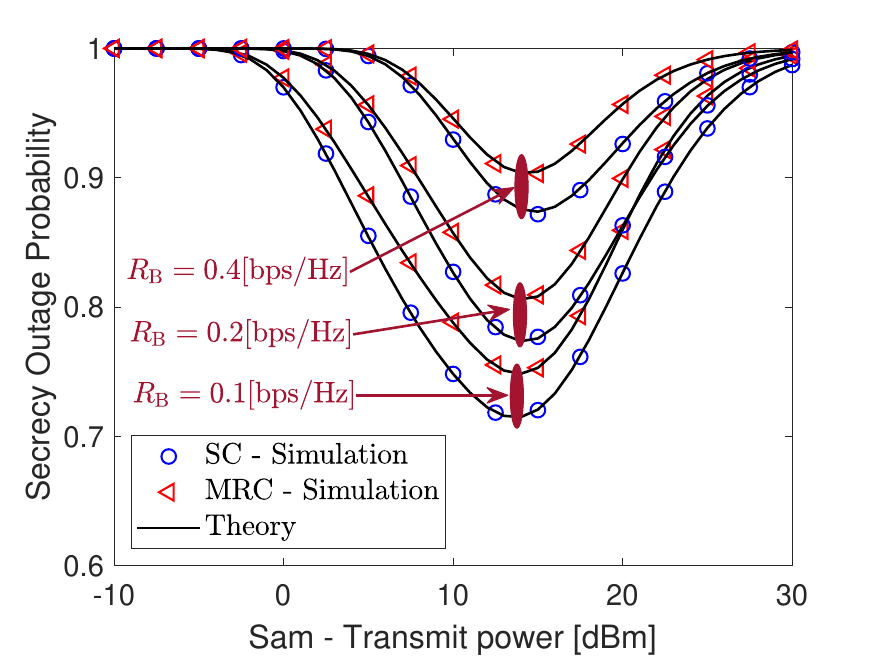}
    \par \footnotesize{b) SOP vs. \(P_s\) at \(\alpha_\bob = \beta_\bob = 0.8\).}
     \caption{SOP performance versus different PA and secret rate configurations. \label{fig:3}}
\end{figure}

Fig.~\ref{fig:3} illustrates the SOP performance as a function of Sam's transmit power $P_s$.
In general, regardless of whether SC or MRC is employed at Eve, the SOP exhibits a convex downward trend.
Notably, the scenario in which Eve uses SC demonstrates better SOP performance compared to the MRC case. This is because MRC offers greater spatial diversity than SC, thereby enhancing Eve’s eavesdropping capability and consequently increasing the SOP.
Fig.~\ref{fig:3}(a) compares the SOP performance for two configurations
of $(\alpha_\will = \beta_\tom = 0.6)$ and $(\alpha_\will = \beta_\tom = 0.8)$.
As observed, increasing $\{\alpha_\will, \beta_\tom\}$, which corresponds to allocating more power to the desired signals in \eqref{Roy-SNR-1-1} and \eqref{Bob-SNR-1-1}, significantly improves SOP performance.
Meanwhile, Fig.~\ref{fig:3}(b) illustrates the effect of the secrecy rate threshold $R_\bob$ on SOP performance.
Intuitively, as suggested by \eqref{eq:SOP}, increasing $R_\bob$ raises the probability $\Pr\left(\sec \leq R_\bob \right)$, thereby increasing the SOP.
This theoretical expectation is consistent with the observed trend in Fig.~\ref{fig:3}(b).
Most importantly, the simulation results align with all analytical findings, thereby validating the accuracy of our SOP analysis.

\subsection{Covert Rate Evaluation}

\begin{figure}[!t]
     \centering
     \includegraphics[width=.85\linewidth]{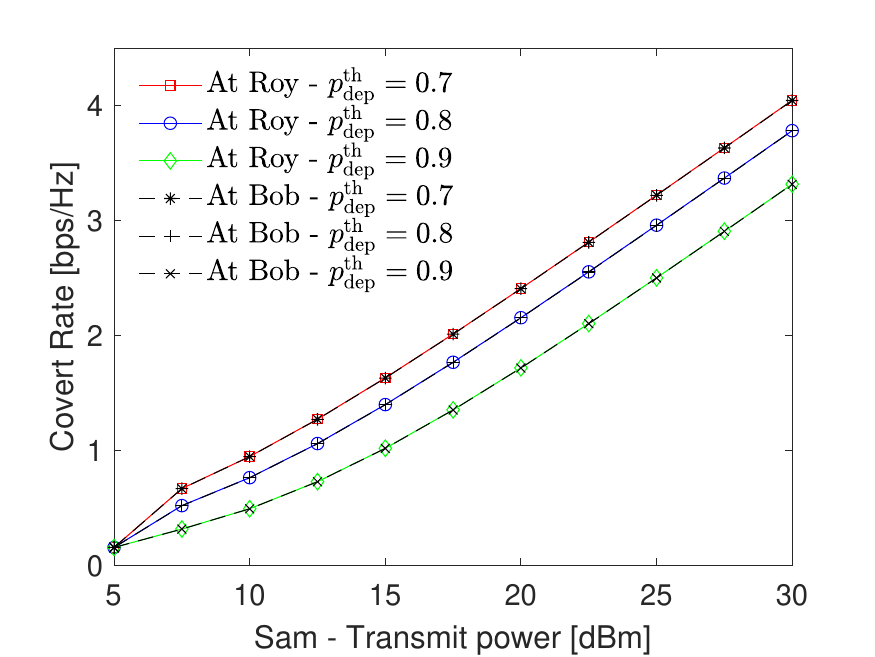}
      \par \footnotesize{a) Covert rate vs. \(P_s\) with \(N=3\) and \(r_\tom=r_\will=0.25\) bps/Hz.}\par 
      
     \includegraphics[width=.85\linewidth]{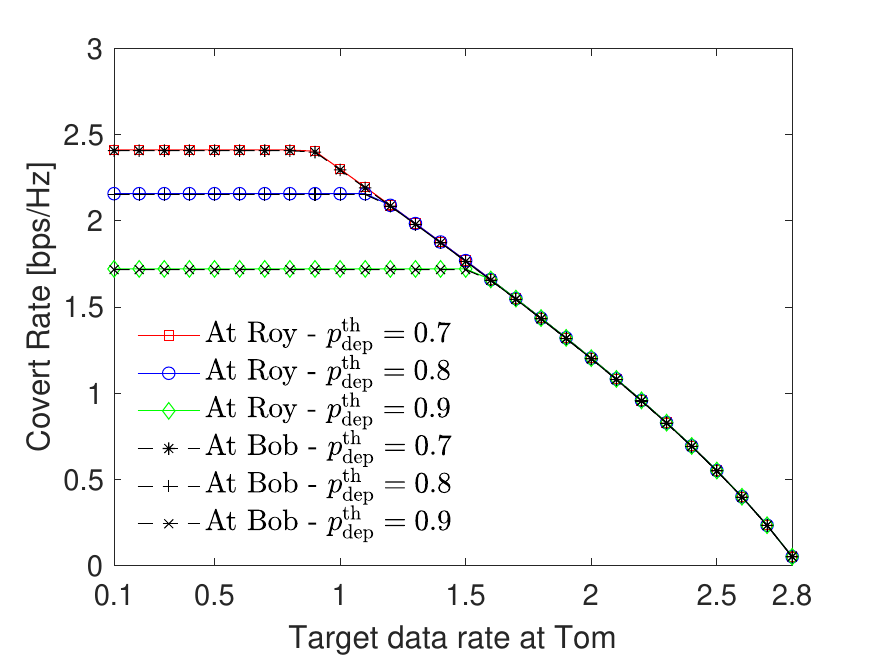}
     \par \footnotesize{b) Covert rate vs. \(r_\tom\) with \(N=3\), \(P_s=20\) dBm, and \(r_\will=0.25\) bps/Hz. }\par     
     
     \includegraphics[width=.85\linewidth]{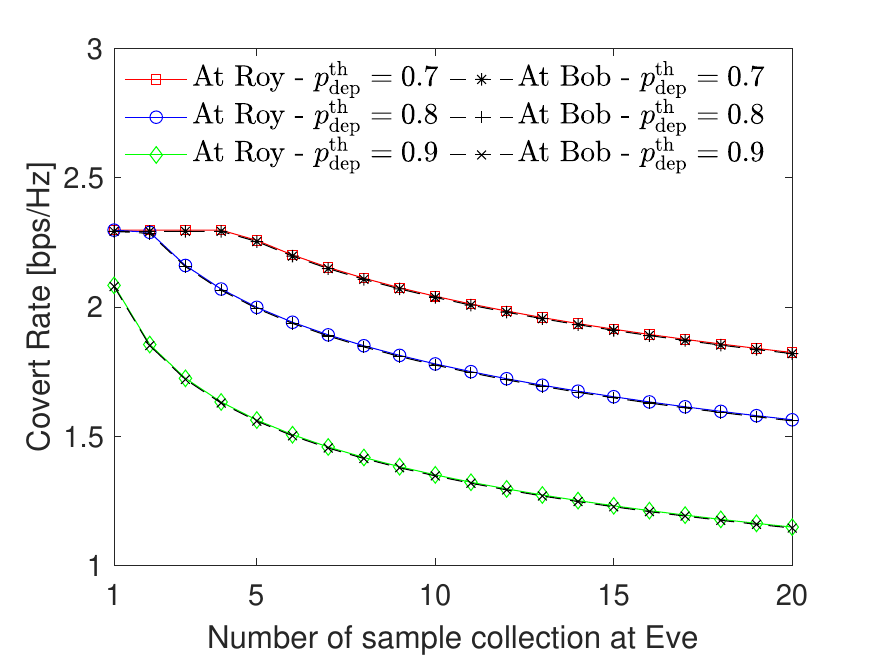}
     \par \footnotesize{c) Covert rate vs. \(N\) with \(P_s=20\) dBm, \(r_\tom=0.5\) bps/Hz, and \(r_\will=0.25\) bps/Hz.}\par    
      
     \caption{Covert rate performance with \(\widehat{r}_\will =0.5\) bps/Hz and \(P_r\) being employed as \eqref{rate-phase-2}.
\label{fig:4}}
\end{figure}

Fig.~\ref{fig:4} illustrates the covert rate performance when the optimal solutions \eqref{alpha-opt-solution} and \eqref{beta-opt-solution} are applied.
The adoption of the optimal values $\alpha_\will^\star$ and $\beta_{\tom}^\star$ ensures that the max-min fairness criterion of the optimization problem in \eqref{problem-2} is satisfied.
This aligns with the results shown in Fig.~\ref{fig:4}, where the covert rates for both Bob and Roy are observed to be identical.
Moreover, from the perspective of the systems' minimum DEP requirements, it is evident that the covert rate decreases as \(\dep^{\rm th}\) increases.
This occurs because a larger value of \(\dep^{\rm th}\) restricts the operational regions of \(\alpha_\will\) in \eqref{eq:theo1} and \(\beta_\tom\) in \eqref{eq:theo2}, resulting in smaller values for \(\alpha_\bob\) and \(\beta_\bob\), and consequently reducing the power allocated to the covert signal \(c_\bob\). 
Fig.~\ref{fig:4}(a) demonstrates that higher values of \(P_s\) lead to increased covert transmission rates of \(c_\bob\) received at both Bob and Roy.
In contrast, Fig.~\ref{fig:4}(b) reveals a degradation in the covert rate as the rate threshold at Tom, $r_\tom$, increases.
This trend is predictable.
Intuitively, a higher $r_\tom$ necessitates an increase in the data rate for decoding $s_\tom$, i.e., $\min\{ \calR_{\bob}^{s_\tom},\calR_{\tom}^{s_\tom}\}$.
Consequently, a higher PA level for the public signal $s_\tom$ (i.e., $\beta_\tom$) is required to satisfy the constraint \eqref{problem-2c}, which in turn reduces the PA level for covert signal $c_\bob$ (i.e., $\beta_\bob$).
As a result, the covert rate $\min\{\calR_{\relay}^{c_\bob},\calR_{\bob}^{c_\bob}\}$ decreases accordingly.
Finally, Fig.~\ref{fig:4}(c) illustrates the degradation of the cover rate as the number of Eve's monitoring samples increases, which is expected due to the enhanced classification capability of Eve with more extensive sampling.

\subsection{Secrecy Rate Evaluation}

\begin{figure}[!t]
     \centering
     \includegraphics[width=.85\linewidth]{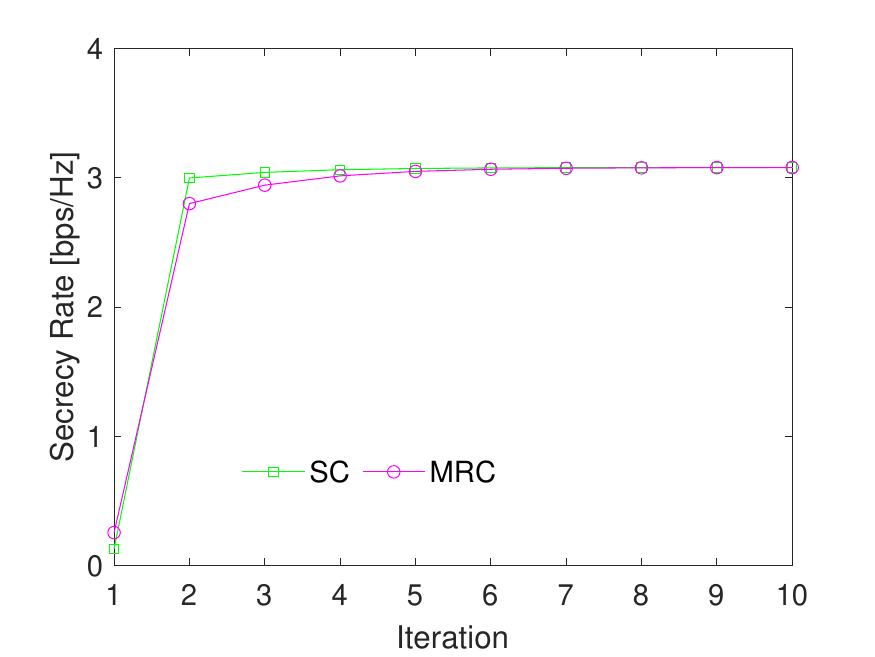}
     \par \footnotesize{a) Convergence behavior of the secrecy rate.}\par

     \includegraphics[width=.85\linewidth]{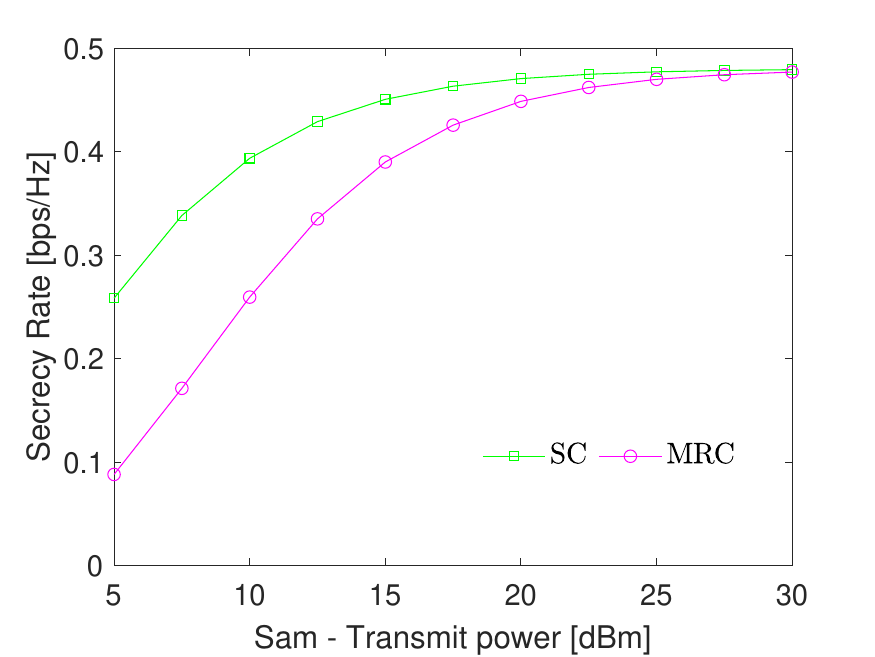}
     \par \footnotesize{b) Secrecy rate vs. Sam's transmit power \(P_s\).}\par

     \caption{Convergence behavior and performance of secrecy rate   with \(r_\tom=\widehat{r}_\will ={r}_\will =0.25\) bps/Hz and \(P_r\) being employed as \eqref{rate-phase-2}.
\label{fig:5}}
\end{figure}

\begin{figure}[!t]
     \centering
     
     \includegraphics[width=.85\linewidth]{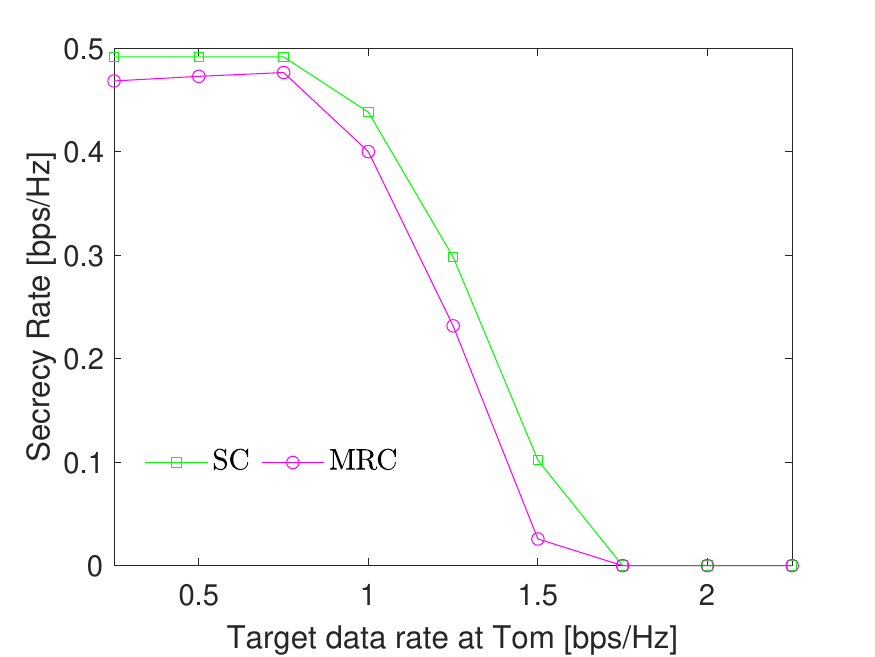}
     \par \footnotesize{a) Secrecy rate vs. \(r_\tom\).}\par

     \includegraphics[width=.85\linewidth]{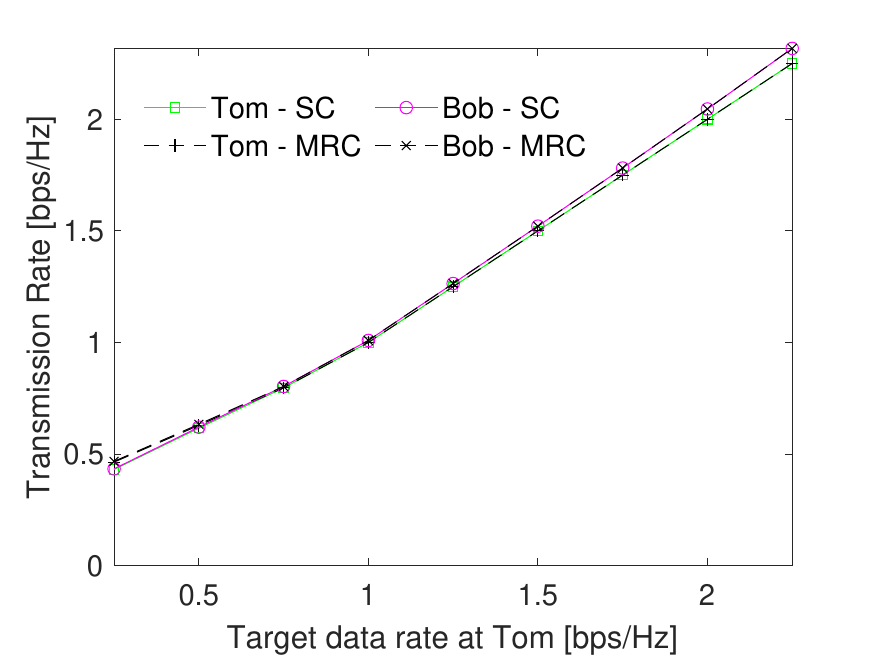}
     \par \footnotesize{b) Transmission rate vs. \(r_\tom\).}\par
     
     \caption{Secrecy rate and transmission rate vs. Tom's outage rate threshold \(r_\tom\), where \(\widehat{r}_\will ={r}_\will =0.25\) bps/Hz and \(P_s = 20\) dBm are assumed, while \(P_r\) is employed as \eqref{rate-phase-2}.
\label{fig:6}}
\end{figure}

Fig.~\ref{fig:5}(a) presents the convergence and feasibility of Algorithm~\ref{Algorith1} for both the SC and MRC scenarios.
It can be observed that the objective function (i.e., the secrecy rate) increases with the number of iterations and converges after approximately five iterations. 
Fig.~\ref{fig:5}(b) depicts the secrecy rate as a function of \(P_s\). 
As shown, the secrecy rate increases with \(P_s\) in the low and moderate regimes and then saturates at around 25~dBm. This behavior arises because, in the high-SNR regime, the secrecy rate ${{\cal R}_s} = {\left[ {{{\cal R}_{\rm{B}}} - {{\cal R}_{\rm{E}}}} \right]^ + } \approx 0.5{\log _2}\left( {\min \left\{ {{\gamma _{\rm{R}}},{\gamma _{\rm{B}}}} \right\}/{\gamma _{\rm{E}}}} \right)$ converges to a constant, as the ratio inside the logarithm function stabilizes.
Furthermore, the secrecy rate achieved by SC exceeds that of MRC in the low and moderate $P_s$ regimes, while both schemes yield identical secrecy rates in the high $P_s$ regime.
Consistent with the discussion in Fig.~\ref{fig:3}, when the MRC scheme provides greater spatial diversity than SC at low and moderate transmit power levels, Eve’s eavesdropping capability is also enhanced under MRC, resulting in a lower secrecy rate compared to SC.
However, in the high $P_s$ region, where $\gamma_{\eva}^{c_\bob}[2]$ approaches a constant, both the SC and MRC schemes become dominated by $\gamma_{\eva}^{c_\bob}[1]$, ultimately leading to identical secrecy rate performance.

Fig.~\ref{fig:6}(a) illustrates the secrecy rate under varying QoS requirements of Tom.
As observed, the secrecy rate remains nearly constant when \(r_\tom\) is less than \(0.75\) bps/Hz. However, once \(r_\tom\) exceeds \(0.75\) bps/Hz, the secrecy rate steadily decreases until approaching zero. 
In contrast, Fig.~\ref{fig:6}(b) shows an increasing trend in the transmission rate with respect to $r_\tom$.
This behavior can be attributed to the same reasoning discussed in Fig.~\ref{fig:4}(b).
Specifically, under the optimization problem \eqref{problem-1}, the constraint \eqref{problem-1c} mandates that a higher $r_\tom$ results in higher transmission rates for decoding $s_\tom$ at Bob and Tom, which aligns with the performance trend observed in Fig.~\ref{fig:6}(b).
This is equivalent to requiring a larger value of $\beta_\tom$ to satisfy the constraint \eqref{problem-1c}, which consequently reduces the value of $\beta_\bob$.
As a result, $\calR_\bob$ decreases, thereby reducing the secrecy rate $\calR_s$, consistent with the trend in Fig.~\ref{fig:6}(a).
Furthermore, it is worth mentioning that the transmission rates for both Tom and Bob meet the rate threshold $r_\tom$ under both SC and MRC scenarios, as evident in Fig.~\ref{fig:6}(b).
\section{Conclusion}\label{sec8}

\subsection{Concluding Remarks}
This study investigates the covertness and security performance of communication between a cellular source and cell-edge users, supported by an IoT master node using the CDRT strategy combined NOMA transmission. 
Considering the effects of Rayleigh fading environments, we derive analytical expressions for the DEP metric across various judgment thresholds and identify the optimal threshold that minimizes DEP in the worst-case scenario.
We also define the effective regions for user PA in NOMA and propose an adaptive PA scheme to maximize the covert rate while maintaining the QoS requirements of legitimate users and guaranteeing the required level of covertness and effective SIC. 
Regarding eavesdropping, we derive expressions for the SOP under two eavesdropping scenarios employing SC and MRC techniques. 
Furthermore, we develop an adaptive PA scheme to enhance the secrecy rate while ensuring the QoS requirements of legitimate users and SIC integrity. Numerical results demonstrate strong consistency between the analytical findings and simulation outcomes and highlight the efficacy of tuning PA coefficients to maximize either the covert or secrecy rate.

\subsection{Open Challenges and Potential Future Research}

In addition to its technical contributions, this work highlights several open challenges and promising research directions for future exploration. For instance, extending the network to multi-antenna systems could offer significant performance improvements. However, this expansion also introduces challenges for energy-constrained wireless IoT devices, which requires a more conservative approach to design and analysis. Another important area for further investigation is the impact of imperfect CSI, SIC, and hardware impairments, in order to better capture the characteristics of practical systems.
Furthermore, beyond the max-min fairness of the covert rate, exploring the fairness of the SOP and secrecy rate between cell-edge and cell-center users in massive connectivity scenarios represents a valuable direction for future research.

\begin{appendices}
\renewcommand{\thesectiondis}[2]{\Alph{section}:}
\renewcommand{\theequation}{\thesection.\arabic{equation}}

\section{Proof of Theorem~\ref{theo1}}\label{AppendixA}
\setcounter{equation}{0}
To prove Theorem~\ref{theo1}, we begin by invoking \cite[Eq. (8.354.2)]{Gradshteyn2014} to express the upper incomplete gamma function as \(\Gamma\left( N, x  \right) = \Gamma\left( N\right) - \sum_{n=0}^\infty(-1)^n \frac{x^{n+N}}{n!(n+N)}\). Accordingly, the optimal DEP in \eqref{dep:close-opt-1} can be rewritten as
     \begin{align}\label{eq:A1}
    \dep^{\star}[1] = 1 -\sum_{n=0}^\infty \frac{(-1)^n\big( 1 - \alpha_\will^{n+N} \big)}{\Gamma(N)n!(n+N)}\left( \frac{N\ln(1/\alpha_\will)}{1-\alpha_\will} \right)^{n+N} .
\end{align}
Since \(\alpha_\will > \alpha_\bob\) and \(\alpha_\will + \alpha_\bob =1\), the operating range of \(\alpha_\will\) lies in \([0.5,1]\).
It is readily obtained that $\partial\dep^{\star}[1]/\partial \alpha_\will >0$, for $ \alpha_\will \in [0.5,1]$.
As a result, $\dep^{\star}[1]$ is a monotonically increasing function with respect to $\alpha_\will \in [0.5,1]$.
Furthermore, as \(\alpha_\will \to 0.5\), \eqref{eq:A1} becomes
 \begin{align}
 \label{dep:close-opt-1-1}
    \dep^{\star}[1] \overset{\alpha_\will \to 0.5}{=} 1 &- \sum_{n=0}^\infty \frac{(-1)^n\left( 1 - (0.5)^{n+N} \right)}{\Gamma(N)n!(n+N)}\nonumber\\
    &\times\left( 2 N\ln(2) \right)^{n+N} < 1 .
\end{align}
For \eqref{eq:A1} when \(\alpha_\will \to 1\), we invoke the limit law \(\lim\limits_{x\to c}f(x) g(x)= \lim\limits_{x\to c}f(x) \lim\limits_{x\to c} g(x)\), where we set \(f(x) =   1 - x^{n+N} \) and \( g(x) = \ln(1/x)/(1-x) \).
Then, we obtain
$\lim_{x\to c}f(x) = 1 - c^{n+N}$ and
$\lim_{x\to c}g(x) =\lim_{x\to c}\frac{\ln(x)}{x-1} = \lim_{x\to c}\frac{1}{x} = \frac{1}{c}$.
By mapping \(x = \alpha_\will\) and \(c=1\), we have 
 \begin{align}
 \label{dep:close-opt-1-2}
    \dep^{\star}[1] \overset{\alpha_\will \to 1}&{=} 1 - \sum_{n=0}^\infty \frac{(-1)^n N^{n+N}}{\Gamma(N)n!(n+N)}\nonumber\\
    &\hspace{-0.25cm}\times  \lim_{\alpha_\will \to 1} \left( 1 - \alpha_\will^{n+N} \right)\left( \frac{\ln(1/\alpha_\will)}{1-\alpha_\will} \right)^{n+N} = 1.
\end{align}  
From \eqref{dep:close-opt-1-1} and \eqref{dep:close-opt-1-2}, it is evident that increasing \(\alpha_\will\) from \(0.5\) to \(1\) lead to the increment of \(\dep^{\star}[1]\) until achieving the peak value of \(\dep^{\star}[1] = 1\). 
When  \(\alpha_\will^{\#}\) is the root of \(\dep^{\star}[1] = \dep^{\rm th}\), the effective operating region of \(\alpha_\will\) becomes \eqref{eq:theo1}. 

\section{Proof of \eqref{CDF-chi-x-close}}\label{AppendixB}
\setcounter{equation}{0}
We begin this proof by first deriving the PDF of \(\chi(x)\) as
\begin{align}
    \label{eq:B1}
   f_{\chi(x)}(z)    
    &= \int_0^z f_{P_s\tau_{\src\eva}}(y)f_{x P_r \tau_{\relay\eva}}(z-y)dy\nonumber\\
    &= \int_0^z \frac{f_{\tau_{\src\eva}}(y/P_s) }{P_s x P_r}f_{\tau_{\relay\eva}}\left(\frac{z-y}{x P_r} \right)dy.
\end{align}
Notably, $\tau_{\cal{X}}$, ${\cal X} \in \left\{ {{\rm{SE}},{\rm{RE}}} \right\}$, is a chi-square random variable, whose its PDF is given by ${f_{{\tau _{\cal X}}}}(x) = \frac{{{x^{N - 1}}\exp \left( { - x/{\lambda _{\cal X}}} \right)}}{{\Gamma \left( N \right)\lambda _{\cal X}^N}}$ \cite{tu2023short,tu2024irs,tu2024multihop}.
By substituting \(f_{\tau_{\src\eva}}(\bullet)\) and \(f_{\tau_{\relay\eva}}\left(\bullet\right)\) into \eqref{eq:B1} and invoking \cite[Eq. (3.383.1)]{Gradshteyn2014}, we get
\begin{align}
   f_{\chi(x)} (z) =  &\frac{z^{2N  - 1}}{\Gamma(2N)(\rho_s x P_r\lambda_{\src\eva} \lambda_{\relay\eva})^{N}} \exp \left( - \frac{z}{x P_r \lambda_{\relay\eva}}\right) \nonumber \\
   &\times  {_1{\calF}_1}\left(N;2N;\frac{P_s \lambda_{\src\eva}  - x P_r\lambda_{\relay\eva} }{P_s x P_r\lambda_{\src\eva} \lambda_{\relay\eva}} z \right).
\end{align}
Accordingly, the CDF of \(\chi(x)\) is computed as 
\begin{align}\label{eq:B3}
    F_{\chi(x)}(z) 
    &= \int_0^z f_{\chi(x)}(z)dz = \int_0^\infty H(1-z) f_{\chi(x)}(z)dz \nonumber\\
    &= \frac{1}{(xP_sP_r \lambda_{\src\eva}\lambda_{\relay\eva})^{N}} \int_0^\infty \frac{ z^{2N  - 1}}{\Gamma(N)}
     \calH^{1,0}_{1,1} \left[ z\left| \begin{matrix}
       (1,1) \\  (0,1)   \end{matrix}  \right. \right] \nonumber   \\
   & \quad\times\calH^{1,1}_{1,2} \left[ \frac{xP_r\lambda_{\relay\eva}- P_s \lambda_{\src\eva} }{xP_r\lambda_{\src\eva} P_s \lambda_{\relay\eva} } z \left| \begin{matrix}
        (1-N,1) \\
        (0;1-2N,1)
    \end{matrix}  \right. \right]  \nonumber \\
    & \quad \times \calH^{1,0}_{0,1} \left[ \frac{z}{xP_r\lambda_{\relay\eva}}\left| \begin{matrix}
       -  \\
       (0,1)
    \end{matrix}  \right. \right]dz,
\end{align}
where the last step is derived by using \(\exp(-z) = \calEFoxH, H(1-z) = \calHFoxH\), and \({_1{\calF}_1}(a,b,z)  = \Gamma(b)/\Gamma(a)  \calFFoxH\) \cite{Prudnikov1986}. Finally, by invoking \cite[Eq. (54)]{TuACCESS2024} to solve the integral in \eqref{eq:B3}, we get \eqref{CDF-chi-x-close}.

\end{appendices}
\bibliographystyle{IEEEtran}
\bibliography{IEEEabrv,reference}
\end{document}